\documentclass[sn-mathphys,Numbered]{sn-jnl}

\usepackage{graphicx}%
\usepackage{multirow}%
\usepackage{amsmath,amssymb,amsfonts}%
\usepackage{amsthm}%
\usepackage{mathrsfs}%
\usepackage[title]{appendix}%
\usepackage{xcolor}%
\usepackage{textcomp}%
\usepackage{manyfoot}%
\usepackage{booktabs}%
\usepackage{algorithm}%
\usepackage{algorithmicx}%
\usepackage{algpseudocode}%
\usepackage{listings}%
\usepackage{xr}
\hypersetup{nolinks=true}

\makeatletter
\newcommand*{\addFileDependency}[1]{
\typeout{(#1)}
\@addtofilelist{#1}
\IfFileExists{#1}{}{\typeout{No file #1.}}
}\makeatother

\newcommand*{\myexternaldocument}[1]{%
\externaldocument{#1}%
\addFileDependency{#1.tex}%
\addFileDependency{#1.aux}%
}

\myexternaldocument{marked_up_main_SI}

\theoremstyle{thmstyleone}%
\theoremstyle{thmstyletwo}%

\theoremstyle{thmstylethree}%

\begin{document}

\title[Prediction of the Cu Oxidation State from EELS and XAS Spectra Using Supervised Machine Learning]{Prediction of the Cu Oxidation State from EELS and XAS Spectra Using Supervised Machine Learning}

\author*[1,2]{\fnm{Samuel P.} \sur{Gleason}}\email{smglsn12@berkeley.edu}

\author[3]{\fnm{Deyu} \sur{Lu}}

\author*[1]{\fnm{Jim} \sur{Ciston}}\email{jciston@lbl.gov}

\affil[1]{\orgdiv{National Center for Electron Microscopy Facility, Molecular Foundry}, \orgname{Lawrence Berkeley National Laboratory}, \orgaddress{\city{Berkeley}, \state{CA}, \country{USA}}}

\affil[2]{\orgdiv{Department of Chemistry}, \orgname{University of California}, \orgaddress{\city{Berkeley}, \postcode{94720}, \state{CA}, \country{USA}}}

\affil[3]{\orgdiv{Center for Functional Nanomaterials}, \orgname{Brookhaven National Laboratory}, \orgaddress{\street{Street}, \city{Upton}, \postcode{100190}, \state{NY}, \country{USA}}}


\abstract{Electron energy loss spectroscopy (EELS) and X-ray absorption spectroscopy (XAS) provide detailed information about bonding, distributions and locations of atoms, and their coordination numbers and oxidation states. However, analysis of XAS/EELS data often relies on matching an unknown experimental sample to a series of simulated or experimental standard samples. This limits analysis throughput and the ability to extract quantitative information from a sample. In this work, we have trained a random forest model capable of predicting the oxidation state of copper based on its L-edge spectrum. Our model attains an $R^2$ score of 0.85 and a root mean square valence error of 0.24 on simulated data. It has also successfully predicted experimental L-edge EELS spectra taken in this work and XAS spectra extracted from the literature. We further demonstrate the utility of this model by predicting simulated and experimental spectra of mixed valence samples generated by this work. This model can be integrated into a real time EELS/XAS analysis pipeline on mixtures of copper containing materials of unknown composition and oxidation state. By expanding the training data, this methodology can be extended to data-driven spectral analysis of a broad range of materials.}

\maketitle


\section*{Introduction}
Due to their wide range of accessible oxidation states and materials applications, the ability to determine the oxidation state of third row transition metals is essential to a wide variety of applications. These include the development of catalysts \cite{Dalle_2019}, photovoltaic devices \cite{Mccusker_2019}, and biotechnology \cite{Johnstone_2015}. Core level spectroscopy is often used to probe transition metal oxidation states, and two main types are electron energy loss spectroscopy (EELS) and X-ray absorption spectroscopy (XAS). EELS provides  detailed atomic scale information, such as oxidation state, coordination number and local symmetry of a nanomaterial \cite{Batson_1993, Browning_1993}. When probing nanomaterials, EELS is often combined with scanning transmission electron microscopy (STEM). In STEM-EELS, an electron beam is scanned over an area of a sample and a full spectrum is acquired and stored at each probe position. This technique is particularly valuable in the study of nanomaterials due to its combination of high spacial and high energy resolution. \cite{Yang_2014, Gazquez_2017, Kociak_2014}. Like EELS, XAS has also attained wide usage in determining oxidation state and local environment in nanomaterials \cite{Bai_2022, Kubin_2018, Henderson_2014}. XAS, however, is typically limited to a spacial resolution of a few nanometers \cite{Yu_2021}, rather than the sub angstrom spacial resolution possible with STEM-EELS \cite{Batson_2002}. The main advantages of XAS compared to EELS for core-loss spectroscopy are the ability to attain higher signal to noise ratios (SNR) and higher energy resolution, particularly at higher excitation energies \cite{Hart_2023}, and functionality on thicker samples for hard x-ray excitation \cite{Akgul_2014}. Due to the myriad use cases for both techniques, they are commonly applied to the nanoscale study of materials containing third row transition metals. However, since EELS and XAS spectra encode the electronic properties of the sample in an abstract way, extracting physical descriptors is a non-trivial task in spectral analysis. 

\par
Therefore, quantitative spectral analysis is often the rate limiting step in materials characterization, and can typically only be conducted by trained experts. This is especially true of L-edge spectra of transition metals, where variations in oxidation state can manifest in small shifts in edge location, L$_2$/L$_3$ ratio and peak width that are not immediately obvious to a non expert, particularly for samples containing multiple oxidation states \cite{Keast_2001}. Oxidation state assignment is typically done by mapping the unknown spectrum to known experimental or simulated standards, a process which can be time intensive and requires significant domain knowledge. Particularly challenging to analyze are mixed valence materials, which are often interpreted as combinations of spectra of integer valence structures \cite{Cressey_1993}. The prevailing solution to this problem is to fit integer valence spectra to the unknown spectrum using least squares. This allows a user to input known standards and determine the coefficients of a linear combination of the standard spectra that reproduce the experimental spectrum \cite{Zhang_2010, Van_Aken_2002, Cressey_1993, Garvie_1998}. 

\par
Least squares fitting has allowed quantitative oxidation state analysis of mixed valence samples, and is widely implemented as the state-of-the-art procedure for quantitative analysis of unlabeled XAS/EELS L-edge data. However, in the case of experimental standards, it has a few serious limitations. First, this procedure requires fresh standards to be taken for each instrument, and often each day, as changes in detector setup and alignment can lead to non trivial changes in the spectra. Second, this procedure is highly sensitive to experimental variation in the acquisition of the standard samples. Contamination with materials of other oxidation state, surface oxidation and beam damage can have a significant impact on the shape of the standard spectrum, and therefore interfere with the fitting of the unknown sample. Additionally, inconsistencies in standard spectrum processing, such as baseline subtraction or the incomplete deconvolution of multiple scattering from the standard sample, can have a similar impact. Third, the presence of any oxidation state or coordination environment unaccounted for by the standards will not only be missed by the prediction of the makeup of the material, potentially missing an important fundamental discovery, but will also lead to an inaccurate representation of the oxidation state as the standard components are forced to represent a signal not originating from any of them. In a similar vein, experimental standards must be taken for every material expected to be present in order to perform the oxidation state analysis. For example, a standard for CuO may not be suitable for an experiment involving CuS due to non trivial differences between the spectra, although they are both a Cu(II) oxidation state \cite{xas_paper, CuS_L_edge_reference}. Simulated standards suffer from fewer experimental limitations, but instead are limited by the level of approximations used in the theory and often can not perfectly reproduce experimental spectra. This can cause systematic errors leading to significant misidentifications, particularly when applied to noisy experimental spectra or experimental spectra more challenging to simulate. It is rare for simulated standards for L-edge transition metal spectra to be quantitatively accurate enough to fit an unknown experimental spectrum using least squares fitting \cite{Groot_2005}. Instead, these are used to qualitatively match components of an unknown spectrum. Therefore, there is a need for a procedure that can determine oxidation state from XAS/EELS L-edge data that is more robust than the least squares fitting of a handful of standard spectra.

\par
An avenue for a more broadly applicable automated analysis procedure is machine learning (ML). Despite some recent advancements in automated L-edge XAS/EELS analysis of transition metals using ML approaches \cite{Timoshenko_2017}, overall, the transition metal K-edge has received more focus from the ML community \cite{Carbone_2019, Torrisi_2020, Zheng_2018}. Numerical analysis of L-edge transition metal XAS/EELS data has mainly been performed using principle component analysis (PCA) to reduce the dimensionality of the spectrum. This field has been well developed, comprising numerous applications of PCA on L-edge XAS/EELS data \cite{Bonnet_1998, Bonnet_2000, Bosman_2006, Kotula_2006}. Additionally, PCA dimenionality reduction procedures have been used to successfully de-noise low SNR core loss EELS data \cite{Cueva_2012, Bonnet_2005, Bosman_2007}. PCA has also been extended into analysis of oxidation states. Applying component analysis to a mixed valence XAS/EELS spectrum can result in components that mimic the unique oxidation states present. This can be used as a qualitative estimation of the different oxidation states present in a sample, however, it is difficult to ensure each of the resulting components match the pure form of an oxidation state. Therefore, the lack of rigorous physical interpretation of the components makes any quantitative analysis challenging \cite{mcr_componnent_analysis_xas}.

\par
Supervised  machine learning approaches have found success predicting oxidation states in manganese and iron samples, using neural networks and support vector machines \cite{Chatzidakis_2019, MnEdgeNet, del_Pozo_Bueno_2023}. However, these models were trained on a small subset of materials and,  with the exception of \cite{MnEdgeNet} on Mn spectra, only focused on integer valence states. Therefore, the more complicated question of L-edge spectra oxidation state regression of an arbitrary Cu  material containing a wide range of oxidation states has not been thoroughly explored. The lack of focus on mixed valence structures generally is especially notable, as such a model is necessary to analyze an in-situ experiment where 1000s of spectra are generated quickly with minor variations in oxidation state.  

\begin{figure}[h]%
\centering
\includegraphics[width=\textwidth]{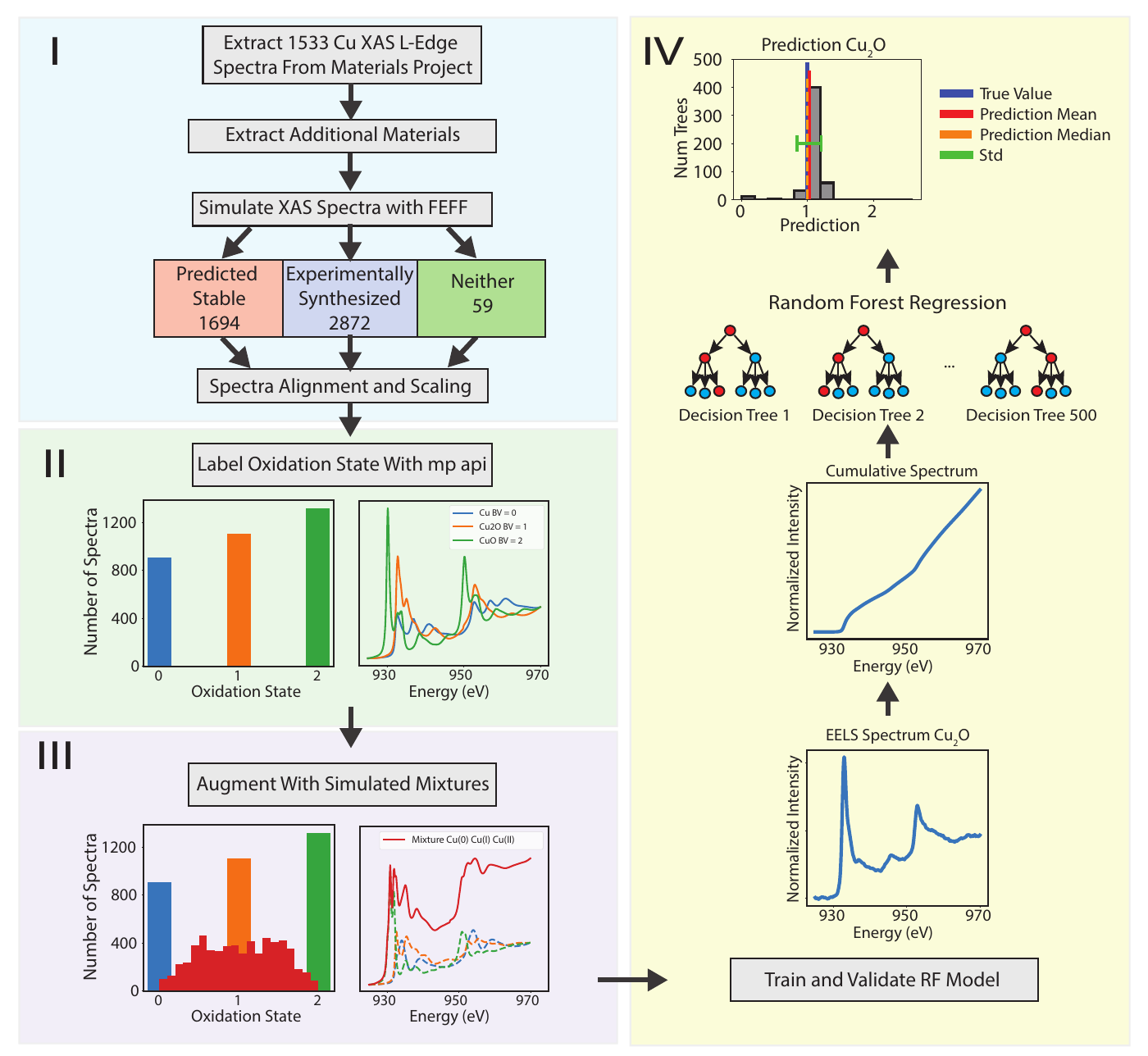}
\caption{A flow chart containing the four components of constructing the training data and random forest model. I, data is extracted from the Materials Project and scaled, aligned and processed to ensure internal consistency and accuracy to experiments. The colored boxes in I show how the materials project classifies the materials extracted and simulated by this work. II, the spectra are labeled by their oxidation state using the Materials Project oxidation state function ``average oxidation states". III, the dataset is augmented by creating mixture spectra made up of linear combinations of integer valence spectra. IV, the random forest model is trained and validated using test simulated data and experimental reference samples \cite{Jain_2013}.}
\label{fig1} 
\end{figure}

\par
This work has developed a supervised ML model capable of conducting a regression task on an unlabeled Cu L$_{2,3}$-edge XAS/EELS spectrum and predicting the average oxidation state. The L$_{2,3}$-edge was selected as the focus due to the prohibitively high energy of the transition metal K-edge for electron detectors. We utilized the simulated L-edge XAS spectra of transition metals stored in the Materials Project \cite{Chen_2021, Jain_2013} as a seed to construct our training set. Despite the differing physical origins of XAS vs EELS, with XAS caused by excitation from a photon and EELS by an electron, under the long wave-length limit and dipole approximation, both spectroscopic methods involve evaluating the same transition matrix element. Therefore, a model trained on XAS data is able to effectively predict EELS data \cite{Egerton_2011, Moreno_2007} for features where the quadrupole contribution is not significant.

\par 
Cu was selected as the focus of this work due to the myriad applications of Cu nanomaterials. Specifically, Cu nanoparticles (CuNPs) are used in antimicrobial agents \cite{Bhagat_2021}, catalysts \cite{Gawande_2016} and renewable energy devices, particularly the electrochemical reduction of CO$_2$ \cite{Laffont_2006}. Examining the oxidation state of Cu nanomaterials is critical to their function, as CuNP preparation procedures can lead to unintended surface oxidation that disrupts many of their applications \cite{Bhagat_2021}. Additionally, the major trends in Cu L-edge spectra can be captured accurately in Cu metal, Cu$_2$O and CuO using the multiple scattering \footnote{In this case multiple scattering refers to the interference of multiple scattering paths, not to be confused with sequential inelastic events originating from the same excitation source.} method implemented in the FEFF9 code \cite{FEFF9_source, Chen_2021}. Figure \ref{feff_to_lit}a-c shows good agreement in the L$_2$-L$_3$ spacing and well preserved intensity ratios between the L$_2$ and L$_3$ peaks. Fine detail such as the splitting of the L$_3$ peak in Figure \ref{feff_to_lit}a is demonstrated as well. The limitations of this method include the treatment of the partially filled 3$d$ bands in Cu(II), where the many-body effects, such as multiplet effects, require higher levels of theory beyond the mean-field level \cite{FEFF9_source, Groot_2005}. This can produce some spurious artifacts in the simulations, such as the L$_3$ shoulder in the CuO simulation (Figure \ref{feff_to_lit}c) which is not present in the experimental sample. Although the quadrupole contribution can play an important role in pre-edge features, distinct spectral features in the main edge regions are found to be sensitive to the oxidation state from feature importance analysis. Therefore, neglecting the quadrupole contribution will not have a significant impact in this analysis. The overall success of FEFF9 in producing Cu L-edge spectra allows Cu materials to serve as a model system for this type of automated analysis procedure. In this work we present a framework for predicting the Cu oxidation state that can be readily extended to other transition metals by acquiring a volume of corresponding simulated XAS data. 

\begin{figure*}[ht]
    \centering
    \includegraphics[width=\textwidth]{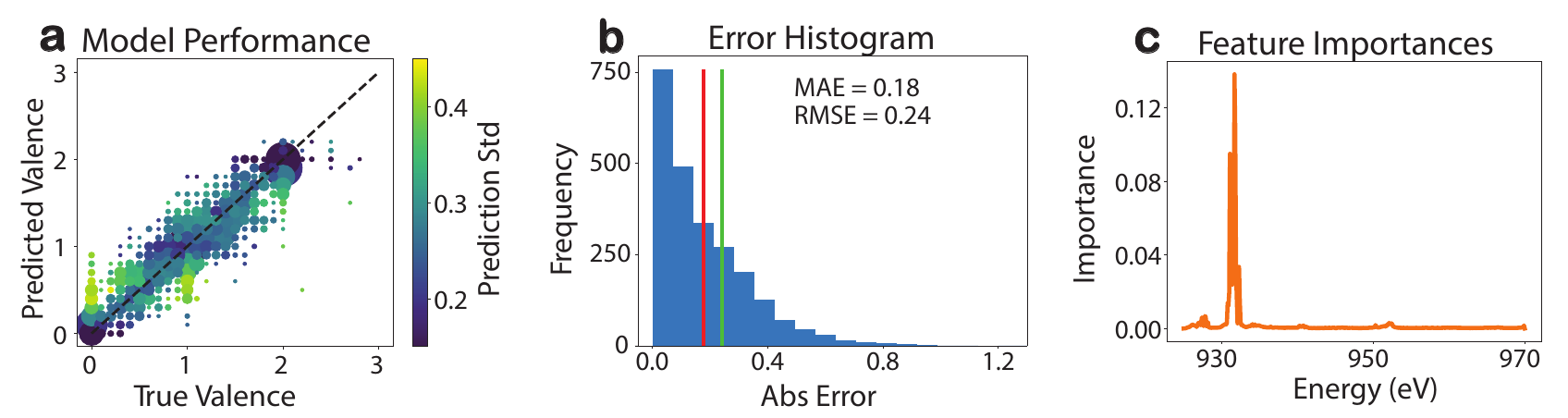}
    \caption{The performance of the random forest model on the test set of simulated data. (a) $R^2$ plot, where each spot's size is proportional to the number of spectra at that point and its color corresponds to the prediction's standard deviation. (b) histogram of the absolute errors, with the vertical green line showing the location of the root mean square error (RMSE) and the vertical red line showing the location of the mean absolute error (MAE).(c) feature importance of the random forest model plotted on the same energy axis as the spectra.}
    \label{fig2} 
\end{figure*}

\section*{Results and Discussion}
\subsection*{Performance on Simulated Spectra}
Our RF model shows a high level of accuracy on a test set of simulated data. Figure 2a shows the $R^2$ plot of the predictions of this test set, which contains roughly 2400 spectra. The $R^2$ for this model is 0.85, and shows a visible high degree of correlation across all the well represented oxidation states. The largest errors come from integer valence misprediction, most commonly when a Cu(0) or Cu(I) spectrum is predicted as mixed valence. However, as shown in Figure 2a, these mispredictions can often be differentiated from the accurate predictions by using the prediction standard deviation (described in the methods section). The feature importance plot from Figure 2c offers insights into the origin of these errors. The model takes a small amount of information from the pre-edge and then bases its prediction mostly on the location and shape of the L$_3$ peak. As Cu(0) and Cu(I) have L$_3$ peaks at almost exactly the same energy, these are harder to differentiate than Cu(II), which is red shifted by roughly 3 eV. Despite this difficulty, Cu(0) and Cu(I) are accurately identified far more often than they are mispredicted, as shown in Figure 2a. As can be seen from Figure 2b, a full integer miss, i.e. a Cu(0) spectrum incorrectly called a Cu(I) spectrum, essentially never occurs. What is even more encouraging in Figure 2a and 2b is the simulated mixture samples are frequently predicted with a high degree of accuracy, showing this model has significant potential in predicting mixed valence samples. 

\subsection*{Model Uncertainty Metric}
In this work we have developed a method for quantifying the uncertainty in our RF model's prediction. This is done by examining the predictions of each of the 500 decision trees which comprise the random forest as well as the averaged value used as the final prediction. This uncertainty analysis is visualized by generating a prediction histogram, as shown in  Figure 1 (IV) and Figure 3d-3f. Beyond the qualitative spread of predictions shown in the prediction histograms, the uncertainty can be understood quantitatively by calculating the standard deviation of these predictions. This is indicated by the horizontal green line in the prediction histogram plots shown in Figures 1 and 3, and is used here as the RF model's internal uncertainty measurement. To leverage this quantitative uncertainty, the standard deviation can be used to filter out predictions that are highly uncertain, and therefore presumably less accurate. Figure \ref{std_threshold} illustrates this concept, where a standard deviation threshold was imposed, and all predictions with a standard deviation higher than this value were discarded due to their high uncertainty. The standard deviation can be used as a powerful tool in determining significantly inaccurate predictions on simulated data, as can be seen when the threshold is set at 0.35 (red rectangle in Figure \ref{std_threshold}a and \ref{std_threshold}b). When this threshold is used, 15\% of the predictions of our test set are higher than the threshold and discarded (Figure \ref{std_threshold}a). However, imposing this threshold causes the RMSE of the remaining 85\% of our test set to decrease 8\% from the full test set value of 0.24 to 0.22 (Figure \ref{std_threshold}b). Therefore, the 15\% of the test set discarded by this method is comprised of predictions less accurate than average, showcasing the utility of this uncertainty metric in informing the accuracy of the model's predictions for unknown samples.    

\begin{figure*}[ht]
    \centering
    \includegraphics[width=\textwidth]{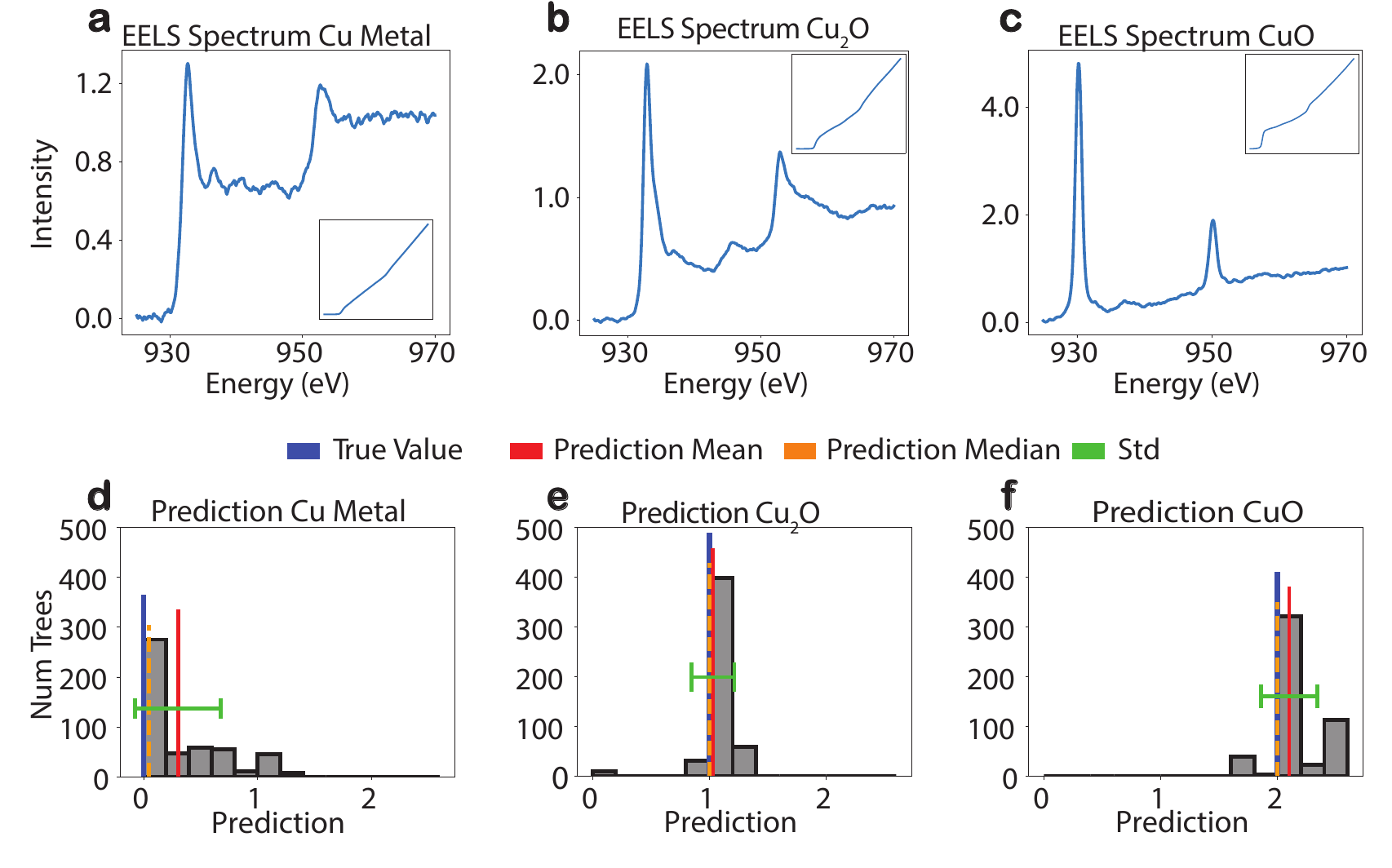}
    \caption{The performance of the random forest model on experimental Cu oxide EELS standards collected in this work. The top row (a, b, c) shows the raw spectrum with the cumulative spectrum as an insert. The bottom row (d, e, f) shows the prediction histograms for each spectrum, where the grey bars correspond to the number of decision trees predicting values over that range, The formal oxidation state is shown by the vertical blue line in each plot, while the predictions generated by the mean and median of the decision tree predictions are shown in red and orange, respectively. The standard deviation of the decision tree’s predictions is shown as a green horizontal line.}
    \label{fig3} 
\end{figure*}

\subsection*{Validation Using Experimental Spectra}
To test the RF model's validity when applied to experiments, we used the model to predict a set of metallic Cu and Cu oxide standards. The simulated spectra corresponding to these standards were left out of both the training and test sets previously discussed. These standards were smoothed using a Savitzy-Golay filter with a window size of 1.5 eV and a polynomial order of 3. From Figure \ref{smoothing_impact} it can be seen that the level of smoothing does not impact prediction accuracy. The smoothing window of 1.5 eV was selected as the default method due to qualitative observations that it removed the vast majority of the noise but also preserved the overall shape of the spectrum (Figure \ref{model_outline}). From Figure 3e and 3f it can be seen that the model has a high degree of accuracy when predicting Cu(I) and Cu(II), rendering essentially perfect predictions for each of these standards, regardless of whether the mean or the median of the decision tree ensemble is used as the prediction. However, Figure 3d shows the Cu(0) standard appears to be slightly over estimated, with the mean prediction rendering a larger overestimate than the median, as the two predict 0.3 and 0.05, respectively. 

There are likely two factors responsible for the overprediction of Cu valence for metallic absorbers. First, as has been discussed above, random forest models average predictions across individual decision trees, in this case 500. Therefore, it will always be more challenging for this model to predict Cu(0) as exactly zero, as all Cu atoms in our training data have non-negative valence. Consequently, any spread in the predictions will result in an overestimate. It is also worth noting that Figure 3d shows that the mode of our prediction histogram contains Cu(0) by a factor of four over the next highest bar, and that the median is much closer to a prediction of Cu(0). A second factor may also partially explain this overestimate, which is that our Cu(0) likely experienced some surface oxidation. Therefore, it may be assumed that this material no longer had a true oxidation state of zero at the time of measurement. This is reflected in the spectrum, which can be seen to have visibly taken on some additional Cu(I) character relative to simulated Cu(0) and Cu(0) observed in XAS studies taken from the literature (Figure \ref{xas_vs_eels}, \cite{xas_paper}). Specifically, our Cu(0) spectrum shows a drop in intensity of the two higher energy peaks in the L$_3$ edge and an increase in intensity in the lowest energy peak, which are characteristic of surface oxidation leading to more visible Cu(I) character. This, combined with the logistics surrounding the attainment of our Cu(0) sample, the sample was not shipped in vacuum sealed vial, and the fact that we were unable to reduce the sample in the microscope, supports the supposition that our Cu(0) EELS sample has undergone some surface oxidation. Therefore, we believe that this prediction of a mixed valence material closer to Cu(0) than Cu(I) matches our experimental realities and a detailed examination of the experimental spectrum. 

However, it is also important to note that an XAS spectrum of Cu(0) extracted from the literature (Figure 4d) is also overestimated by our model. This is unlikely to be a result of surface oxidation, both from an instrumental perspective and a qualitative examination of the spectrum, which shows a much more characteristic Cu(0) sample than our experimental EELS sample due to the relatively equal heights of the three L$_3$ edge peaks and the L$_3$ being lower in intensity than the L$_2$ edge (Figure \ref{feff_to_lit}, S7). We attribute the overestimation of the literature Cu(0) mainly to the fact that our model does not allow for non-positive predictions, which causes any uncertainty in the prediction of Cu(0) to result in an overestimate, as discussed above.

In addition to the experimental spectra predicted here and in the following section, 8 other spectra extracted from the literature were predicted using this model (Figure \ref{other_exp_samples}, \ref{shift_other_exp_samples})\cite{Rudyk2011ElectronicSpectroscopy, Goh2006TheAir, Blanchard2010ElectronicSpectroscopy}. 7 of these 8 spectra were materials with a Cu(I) oxidation state and all are predicted to within 0.1 of Cu(I) when the edge alignment is correct, and most retain their accuracy when the edge is misaligned by 0.5 eV in either direction (Figure \ref{shift_other_exp_samples}). The one Cu(II) material, CuS, is predicted as roughly 1.5, however, our model’s prediction is likely inaccurate due to this spectrum's high intensity post the L$_3$ region (Figure \ref{CuS}). As the Cu(II) L$_3$ edge is roughly 2eV lower in energy than Cu(I) and Cu(0), this increased intensity is likely mimicking a mixed valent spectrum, with this extra intensity appearing to come from absorption from a Cu(I) material. XAS is impacted by multiple photon scattering in very thick samples, which produces artificially high intensity in the tail of the L$_{2,3}$ edge spectrum. Therefore, we believe that this was simply a spectrum of  a very thick sample, leading to multiple scattering induced changes to the spectrum that the model is unable to account for.

\begin{figure*}[ht]
    \centering
    \includegraphics[width=1\textwidth]{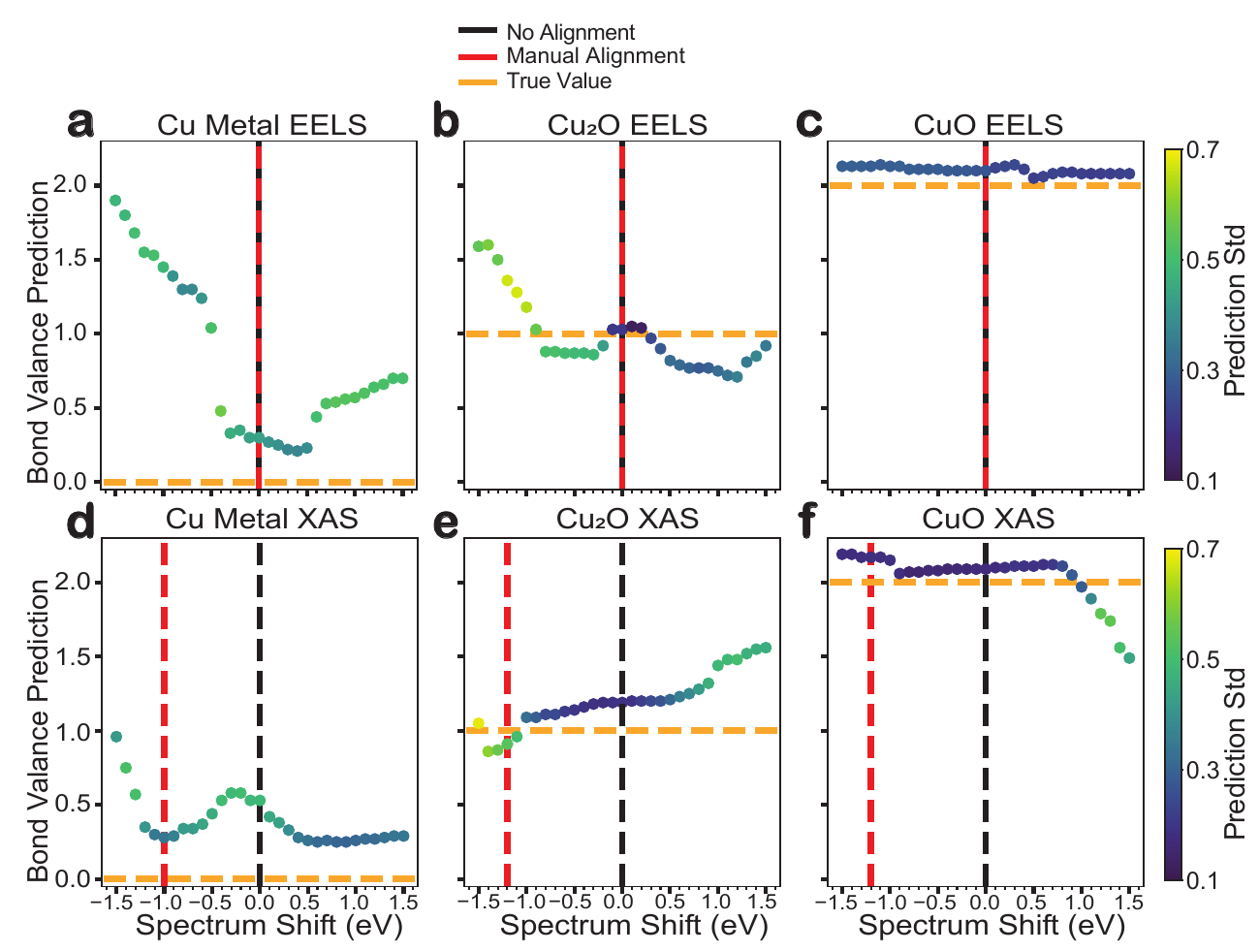}
    \caption{The impacts of energy misalignment on the prediction for EELS spectra taken in this work (a-c) and XAS spectra extracted from the literature (d-f) \cite{xas_paper}. The spectra are shifted horizontally on the energy axis by the amount indicated in the x axis but are not changed in any other way. The scatter plot color corresponds to the prediction's standard deviation. The horizontal orange dashed lines show the location of the formal oxidation state of each material. The dashed black line indicates the location of the raw spectrum without any energy axis shifting. The red dashed line indicates the amount of energy axis shifting required to bring the experimetnal spectrum's onset energy to the same value as its corresponding simulated spectrum in our dataset.}
    \label{fig4} 
\end{figure*}

\subsection*{Energy Axis Misalignment}

\par Given that we have performed a manual edge alignment correction to our training data, we also examine the impact of energy axis misalignment on our predictions of experimental spectra. To explore this, we created a set of experimental spectra where the onset energy was shifted by controlled amounts and tracked how this shift impacted the oxidation state prediction (Figure 4). From Figure 4a we see that the energy misalignment has the greatest impact on the Cu(0) sample, and an offset of -0.4 eV or greatercauses an inflection point where the prediction jumps from 0.3 to nearly 0.5. Misalignment in the positive direction has a far less dramatic impact, and an energy shift of +0.5 eV produces essentially no change in the prediction. The Cu(I) sample, shown in Figure 4b, is more stable, with a shift of nearly 1 eV in either direction resulting in a change of less than 0.2 in the oxidation state prediction. In Figure 4c, we see that Cu(II)'s prediction is virtually independent of shift plus/minus 1 eV, which is likely explained by the greater than 2 eV gap between the onset energy of Cu(II) vs Cu(I) and Cu(0). 

\par To further examine the utility of our model when applied to experimental spectra, and to further study the impact of absolute energy shift, an additional experimental validation was done using an extracted set of XAS spectra of Cu oxides \cite{xas_paper}. This set of spectra has been measured to be shifted from the experimental spectra used to validate this model by -1.0 eV for the Cu metal spectrum and -1.2 eV for the Cu$_2$O and CuO spectra (Figure \ref{feff_to_lit}), and provides a test case for how the model will respond to spectra with their energy axes significantly misaligned. From Figure 4d-f, we can see that our ML model produces excellent results for the XAS spectra when they are correctly aligned to our training data (red line in Figure 4d-f) and the results are robust even when the raw spectra are predicted, which are severely misaligned (black line in Figure 4d-f). When such a misalignment has occurred, the Cu(I) and Cu(II) spectra are predicted with near perfect accuracy, while the Cu(0) spectrum appears to be slightly over estimated, returning a prediction of around 0.5 when the correct alignment prediction is 0.28. It is worth reflecting this prediction is still an overestimate, although closer to zero than our experimental EELS spectrum shown in Figure 3d, reflecting this model's propensity to overestimate Cu(0). With these observations, it is clear that the ML model trained on properly aligned spectra can achieve highly accurate results on spectra with significant energy misalignment. Additionally, a potential avenue to determine the true alignment location is to vary the energy axis and seek out regions of consistent stability and low prediction standard deviation, as these regions are often associated with more accurate predictions for our experimental data.

\begin{figure*}[ht]
    \centering
    \includegraphics[width=1\textwidth]{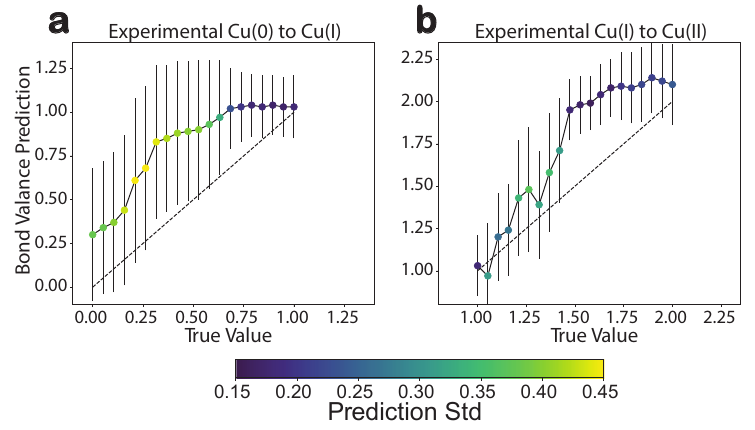}
    \caption{Performance of the random forest model on experimental mixed valence spectra. (a) mixtures of Cu(0) and Cu(I), while (b) shows mixtures of Cu(I) and Cu(II). The scatter plot color corresponds to the prediction's standard deviation. The dashed line indicates the location of a perfect prediction, while the vertical lines indicate the standard deviation of each prediction. The vertical lines show one standard deviation in the negative and positive direction.}
    
    \label{fig5} 
\end{figure*}

\subsection*{Prediction of Experimental Mixed Valence Samples}
Post successful proof of concept for our model on standard experimental samples, we turn our attention to a more valuable, but also more challenging, experimental case, the prediction of samples of mixtures of different oxidation states. As shown in Figure 2a, our model has already demonstrated a high degree of accuracy on simulated mixed valence samples. Additionally, we show how smooth variance in simulated mixed valence materials excluded from the training data is captured by our model by showing simulated mixtures of Cu(0), Cu(I) and Cu(II) in Figure \ref{simulated_mixtures}. The important test for the utility of this model in experimental spectra is how well this process works on experimental mixtures of oxidation states. Due to the difficulty in engineering a system with smoothly varying mixed valence states, and inherent uncertainties in quantifying such a system, we have generated mixed valence experimental spectra through linear combinations of our standard samples. The labeled value for these experimental mixtures is determined by multiplying their formal oxidation state by their contribution to the final mixture spectrum, as was done with the labeling for the simulated mixtures. For example, a mixture of 40\% Cu(0) standard and 60\% Cu$_2$O standard would be calculated as follows: 

\begin{equation}
0.00*0.4 + 1.00*0.6 + 2.00*0.0= 0.6
\end{equation}

The results are shown in Figure 5. From Figure 5a-b, we see both plots contain regions of high accuracy, particularly for mixtures of Cu(I) and Cu(II) (Figure 5b). These mixtures are accurately predicted to within less than 0.1 in close to half of the mixture samples. However, we can see that the absolute accuracy has sections of low accuracy, particularly at inflection points where the prediction is changing quickly. This is particularly true for mixtures of Cu(0) and Cu(I) (Figure 5a), where the inflection region drives the prediction into a region of significant overestimation which is not recovered until the mixture becomes entirely Cu(I). However, the overall trend of the prediction is correct, as in both Figure 5a and b the higher valence sample is identified as such until a pure sample is predicted, regardless of any absolute inaccuracies in the prediction. 

Both mixed valence cases tend overestimate the oxidation state when the higher oxidation state sample comprises greater than 50\% of the mixture. We believe this to be a feature of the higher maximum intensity and sharper peaks of higher oxidation state spectra. This can be seen in Figure 3a-c, where the edge intensity relative to the tail of each spectrum is shown. This is also noticeable in the cumulative spectrum, which is the actual input into the model, (inset in Figure 3a-c) where the lower intensity in the Cu(0) peak results in an almost linear cumulative spectrum, while the higher intensity of Cu(I) and Cu(II) are very noticeable as a sharply increasing region in the L$_3$ edge region of the cumulative spectrum. The higher intensity of the higher oxidation state may make the fine features of Cu(0) difficult to detect at low mixture fractions, as Cu$_2$O and Cu(0) have their onset edges and L$_3$ peaks at essentially the same energy. Additionally, The slight red shift in Cu(II) spectra yields an immediately noticeable feature for model identification, and a sample which is 75\% Cu(II) and 25\% Cu(I) may simply be predicted as a Cu(II) with a shoulder or other unusual transition, which is relatively common in the simulated data. 

We have also predicted random experimental mixtures of Cu(0), Cu$_2$O, and CuO. This was done using mixtures of the literature XAS spectra and our experimental EELS data. The results are shown in Figure \ref{empirical_correction}a and c, respectively. They contain a characteristic overestimation as seen in the smoothly varying mixed valent binary mixtures, which inspired the creation of an empirical correction to random mixed valent spectra. The literature XAS predictions were used to train a linear regression model to predict the true oxidation state based on the prediction of the unknown spectrum. This was done using the mean, median and standard deviation of the decision trees as input. The model was trained and validated using the random mixtures of the literature XAS data and tested on the predictions of the random mixtures of EELS data. Figure \ref{empirical_correction}a-b shows the predictions on the training data, the literature XAS sample, and how the empirical correction improves the predictions. Upon generation of this empirical model, it was used to correct the predictions of mixtures of Cu(0), Cu$_2$O, and CuO experimental EELS spectra. The results of applying this correction are shown in Figure \ref{empirical_correction}c-d. The generation of this empirical model shows that, despite the challenges of predicting a pure Cu(II) in the empirical model’s predictions, the overall trend of the mixed valent overestimation can be captured and corrected.

\subsection*{Comparison of Prediction Methods}
As Figure 3d clearly shows that the median/mode of the decision trees yields a prediction much closer to Cu(0) for our experimental EELS Cu(0) sample, and Figure 5 shows a characteristic overestimation in both types of experimental mixtures, it is worth exploring whether using the median or the mode as a prediction, rather than the mean, is more accurate overall. This idea is explored in detail in Figure \ref{mean_median_and_mode}, which shows the full prediction histogram for various mixtures of Cu(0) and Cu$_2$O (Figure \ref{mean_median_and_mode}a-f) and compares the median, mean and mode predictions directly (Figure \ref{mean_median_and_mode}g-i). From Figure \ref{mean_median_and_mode} we see that, although the median is more accurate for pure Cu(0), it quickly begins to overestimate the oxidation state by a margin significantly greater than the prediction using the mean of the decision trees. The mode prediction is even more extreme, jumping from predicting Cu(0) for pure Cu(0) and 80/20 mixtures of Cu(0) and Cu$_2$O, to a prediction of Cu(I) for 60/40 mixtures of Cu(0) and Cu$_2$O. Therefore, for predictions of mixtures of oxidation states, the mean prediction is more likely to be accurate, and mean/mode statistics can be helpful in instances where a pure oxidation state is assumed, particularly if the material is expected to be metallic. Due to this utility, predictions using this model return the full set of predictions as well as the mode, median and the mean of the decision trees.    

\subsection*{Impact of Noise on Simulated Data}
To test the impact of noise on the simulated data, random Poisson noise was added to each simulated spectrum in the test set to produce a test set augmented by noise. To ensure that this process echoed our approach on experimental spectra as much as possible, the simulated spectra, which are on a 0.1 eV resolution, were re-sampled using scipy's 1d interpolation function with a higher resolution of 0.03 eV, matching that of our experimental samples. Noise was then added to the interpolated spectra, and these spectra were then smoothed in the same method as the experimental spectra and integrated to produce a cumulative spectrum (Figure \ref{model_outline}). These spectra were then predicted by the model to test its accuracy on noisy data.

\begin{figure*}[ht]
    \centering
    \includegraphics[width=1\textwidth]{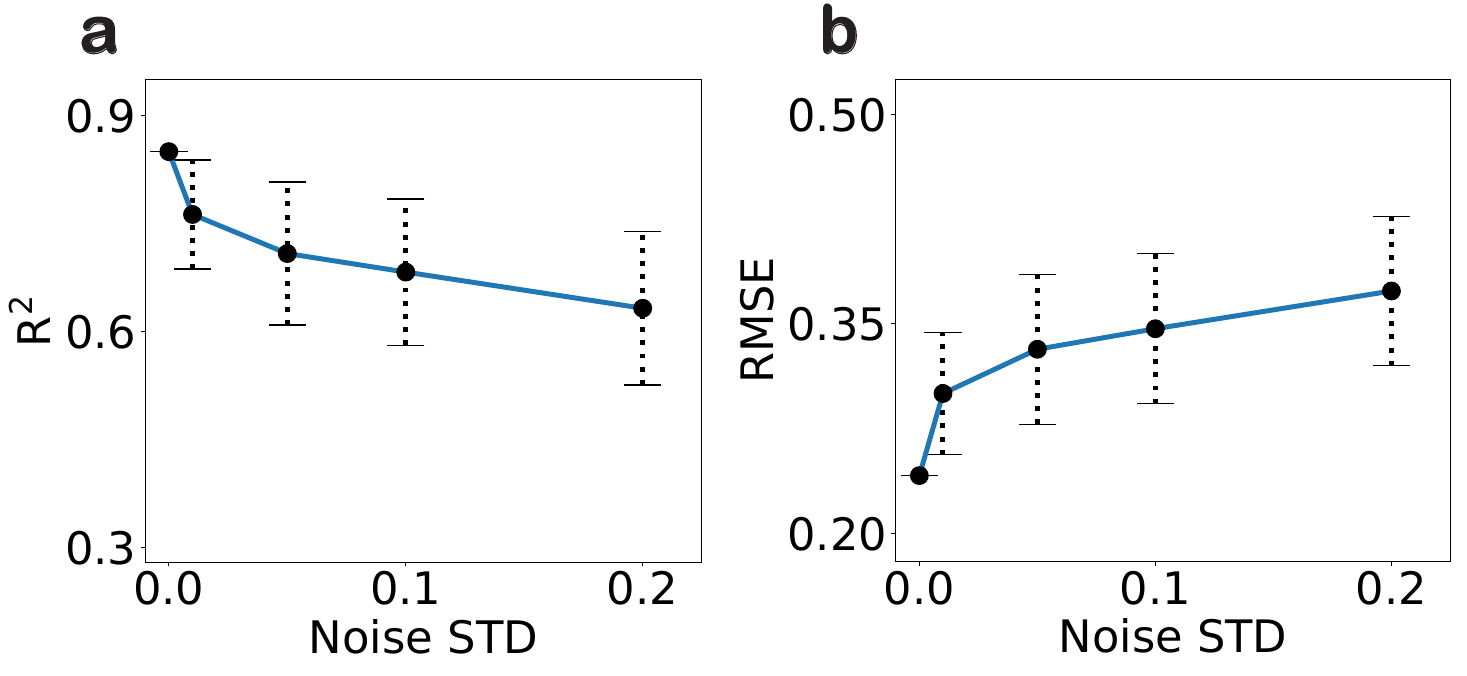}
    \caption{Random forest model performance on simulated data augmented by Poisson noise. The standard deviation of the Poisson distribution used to generate the noise is shown in the x axis of each plot. The error bars denote the standard deviation of the RMSE/R$^2$ across 100 random states for that noise standard deviation value.}
    \label{fig6} 
\end{figure*}

As shown in Figure 6a, the simulated data are relatively sensitive to noise augmentation, and the addition of a small amount of Poisson noise resulted in an increase in RMSE from 0.24 to 0.3 as compared to results from the noiseless spectra. Further increase in noise led to an even larger RMSE, however the decline in accuracy becomes less sharp than the initial slope. A similar trend is seen in Figure 6b for $R^2$, where a drop in $R^2$ is observed after adding a small amount of noise, however this decline is less sharp than the increase in RMSE, and adding additional noise has a more pronounced decline on $R^2$ than subsequent noise does on RMSE. Despite this observation, the noise level of our experimental spectra, which are noticeably larger than the simulated low noise case, do not appear to suffer as much as these simulated noisy spectra (Figure \ref{model_outline}). An examination of the quantitative noise level of the experimental spectra can be found in Figure \ref{quantify_noise}, which shows that the noise STD for the experimental EELS spectra is between 0.03 and 0.05. Additionally, the selection of the random seed for the addition of noise appears to have a significant impact on the overall accuracy of the noisy test set. This is shown with the error bars in Figure 6a and 6b, which represent the standard deviation across 100 different random noise seed states. The presented RMSEs and $R^2$s are the average values across these 100 random states. A detailed examination of the noise profiles for these higher error random states shows that in these spectra the region around the baseline experiences noise spikes that mimic features around the baseline region, similar to how an inaccurate power law subtraction of an EELS spectrum baseline appears. This observation further enforces that the accuracy of this model relies heavily on the accurate identification and subtraction of the baseline.  

\section*{Conclusion}
In this work, we have built a random forest model trained on simulated L-edge XAS spectra which is capable of predicting the oxidation state of copper based on its L-edge XAS/EELS spectrum. We have also developed a database of Cu XAS spectra containing 3500 unique materials that have been accurately aligned to experimental spectra, and augmented this database with 6000 simulated mixture spectra. Our random forest model attains an $R^2$ of 0.85 on simulated data with an RMSE of 0.24 and has been shown to accurately predict experimental spectra taken from our home institution and from the literature. Additionally, this model has proven successful predicting mixed valence samples, showing its applicability to track Cu oxidation state in in-situ experiments where oxidation state is changing fluidly as a reaction occurs. Beyond this model's utility to Cu materials, we have also developed a broader methodology which can be extended to the analysis of other materials by acquiring a spectral database of accurate simulated L-edge spectra for the corresponding material.  

\section*{Methods}

\subsection*{Training Set Generation}
In this work, simulated FEFF9 XAS spectra of Cu materials were extracted from the Materials Project. This initial extraction produced a dataset of site averaged spectra for 1533 materials, which contains the 59 materials shown in Figure 1I labeled as neither predicted stable nor synthesized \cite{Chen_2021}. To increase the volume of our training data, an additional 2000 structures were selected by searching the Materials Project for all Cu containing materials that had either been previously synthesized or were predicted to be stable by theory \cite{Jain_2013}. This choice screens a broad material space that is likely accessible to experiments. We computed 2000 site averaged spectra using the \emph{Lightshow} workflow~\cite{Lightshow} and FEFF9 \cite{FEFF9_source}. The combination of this augmentation step and the initial extraction of L-edge spectra already generated by the Materials Project provided 1199 materials that both have been experimentally synthesized and are predicted to be stable (Figure 1I). For each structure, unique Cu sites are determined by the space group symmetry. Then site specific spectra were calculated using FEFF9. The L$_2$ and L$_3$ spectra for each site were combined into the L$_2$,$_3$ spectrum by summing the L$_2$ and L$_3$ spectra, after first interpolating onto the same energy grid (Figure \ref{L2_3}). The site averaged spectrum is calculated from the weighted sum of site-specific spectra based on the multiplicity of the unique sites in the unit cell. The oxidation state of the site averaged spectra spectra were determined using the Materials Project's ``average oxidation states" function \cite{Jain_2013}. Despite this averaging procedure, greater than 93\% of the site averaged spectra retained integer valence. When FEFF9 failed to converge for some, but not all, of the sites in a material, converged site spectra were averaged leaving out the failed spectra. 

To prepare our training set of 3500 site averaged spectra, several additional steps were performed. This workflow is summarized in Figure 1. First, spectra were interpolated to ensure they were all on a 0.1 eV energy resolution. Second, the non uniformity in the energy range of the L$_3$ edge, specifically at the starting point, was addressed by fitting a 6th order polynomial to connect the lowest energy point to [925, 0] (i.e., vanishing intensity at 925 eV) for every spectrum (see Figure \ref{baseline}). The spectra were then aligned to ensure their onset edges were in the same general energy range as those seen in experimental EELS Cu materials, as it was observed that FEFF9 was producing a systematic misalignment in the absolute energy of the L$_{2,3}$ edge . 

\par To accomplish this alignment, two systematic errors were corrected. First, a high degree of onset energy variability was observed across zero valence materials, which would be expected to all have similar onset energies. Second, the absolute energy of the simulated spectra were several eV off from experimental standards. Both of these issues were fixed simultaneously by our automated alignment procedure. Below is a brief summary of our edge alignment procedure following the $\Delta$SCF method \cite{England2011OnDioxide, Meng2024MulticodeCompounds}, as shown in equation (2):

\begin{equation}
     E_{align} = 
     E_{raw} + (\epsilon_{core} - \epsilon_{Fermi}) + (\epsilon_{XCH} - \epsilon_{GS}) + \Delta
\end{equation}

where $E_{raw}$ and $E_{align}$ are the excitation energies before and after alignment. In order to correct the inaccuracy in the calculated excitation energy, we scale the raw spectrum by the difference between the Fermi energy ($\epsilon_{Fermi}$) and Cu 2p core level ($\epsilon_{Core}$) and by the total energy difference between the core-hole excited state ($\epsilon_{XCH}$) and the ground state ($\epsilon_{GS}$). In the core-hole excited state, the core electron is placed at the bottom of the conduction band, known as the excited core-hole (XCH) method. After this alignment, there is a single empirical constant ($\Delta$) calibrated on a reference system to account for the residual discrepancy between theory and experiment. In our study, $\epsilon_{Fermi}$ is taken from the FEFF9 output corresponding to k=0, where k is the photoelectron wave number. $\epsilon_{Core}$ is set to -916.8226 eV, which is determined by the VASP estimation of the energy of a 2p core hole in Cu\cite{Meng2024MulticodeCompounds, Kresse1996EfficiencySet}. ($\epsilon_{XCH}$ - $\epsilon_{GS}$) was computed using the VASP code base, and the values are -650.888 eV, -650.748 eV and -651.945 eV for Cu, Cu$_2$O and CuO, respectively \cite{Kresse1996EfficiencySet}. In principle, one should perform VASP calculations for all the systems in the database. However, this will lead to a very high computational cost, which is impractical for the scope of this study. Therefore, we treated ($\epsilon_{XCH}$ - $\epsilon_{GS}$) as constant for each oxidation state, using the Cu, Cu$_2$O and CuO values listed above for Cu(0), Cu(I) and Cu(II) spectra respectively. 
This resulted in a simplification of equation (2), where ($\epsilon_{XCH}$ - $\epsilon_{GS}$) + $\Delta$ is treated as a constant, $\delta_{ox}$, with different values for each oxidation state. These are: 1849.06 eV, 1849.33eV and 1846.87 eV for Cu with oxidation state of 0, +1 and +2, respectively, which aligns simulated Cu, Cu$_2$O and CuO spectra to their corresponding EELS experimental spectrum.
For the small subset of materials that were classified as mixed valence, they were aligned based on whichever integer oxidation state they were closest to.

\par It is important to note that this alignment procedure is not aligning the edge to the exact location of the Cu/ Cu$_2$O/CuO edges for all Cu 0, +1 and +2 spectra (i.e. forcing every Cu(II) spectrum to start at 930.2 eV, where the CuO edge is located). This alignment procedure computes a correction based on FEFF9’s Fermi energy prediction and then uses the energy gap between the simulated spectrum of either Cu metal, Cu$_2$O and CuO, post Fermi correction, and the corresponding experimental spectrum to scale all spectra with that oxidation state. For example, not every Cu(II) spectrum is at the same edge energy, and many of them are quite different based on their initial location post Fermi energy correction. The relative energy alignment from FEFF9 within an oxidation state is often preserved, particularly for Cu(I) and Cu(II) materials, and this scaling using the experimental spectra is done to bring the energy axis to experimental relevance. Without this correction, the energy axis for the FEFF9 spectra is misaligned by multiple eV and any experimental prediction is impossible. An example of this alignment procedure is shown in Figure \ref{energy_alignment}.

\par Our spectral dataset was then augmented by generating simulated mixed valence samples (see step III in Figure 1, Figure \ref{mixture_generation}). To accomplish this, 300 random sets of spectra were drawn from the integer dataset, each draw taking a random Cu(0), Cu(I) and Cu(II) site averaged spectrum. Each of these 300 sets of 3 integer spectra were then linearly combined to mimic mixed valence structures. For each set of three spectra, 20 random fractions of each material were combined to produce a simulated mixed valence spectrum. To ensure an even spread of mixed valences, 100 sets were combinations of Cu(0) and Cu(I), 100 were combinations of Cu(I) and Cu(II), and 100 were combinations of Cu(0), Cu(I) and Cu(II). This mixture produced a final dataset of roughly 9500 spectra with data well distributed from Cu(0) to Cu(II) (step III in Figure 1, Figure \ref{mixture_generation}). Our training and test sets were generated by separating classes of mixtures, rather than a random 75/25 split across the full 9500 spectra dataset. To accomplish this, we tracked the compositions of each random mixture, and ensured each composition was fully placed in either the training or test set. For example, an arbitrary  mixture of Cu(0), Cu(I) and Cu(II) would have 20 random proportions of each material in our full dataset, and our train/test split ensured all 20 of these were either in training or test. This ensures the model is not biased by seeing a 0.3 0.3 0.4 mixture of the above materials in training and then tested on a mixture of 0.2 0.3 0.5 of the above compounds, which results in a very similar spectrum.  

\par To achieve the best ML model performance, we have tested different spectral representations, including the spectrum itself, its first and second derivative, and the cumulative integral of the spectrum. We found that the best model performance was achieved with the cumulative integral with intensity normalized to 1. In addition, using the cumulative integral, referred to as a cumulative spectrum in this work, as input feature can ensure consistency in the absolute scale of the EELS spectrum. This representation can simplify intensity scaling, as experimental post processing decisions and noise can create a high degree of variability in spectral intensity. The cumulative spectrum approach is insensitive to the absolute scale of the spectrum, although it does require an accurate identification and subtraction of the baseline for experimental spectra. 

\subsection*{Random Forest Modeling}
Random forest (RF) models for this work were trained using Scikit-learn's RandomForestRegressor model \cite{sklearn_source}. The number of trees was fixed at 500, with all features available and  max depth unfixed. The dataset was split into train and test components using a 75/25 random train test split function from Scikit-learn. The structure of this model allows for the input of a raw spectrum of arbitrary min and max energy and energy scale. The model then takes the input spectrum and interpolates it to a 0.1 eV resolution from 925 to 970 eV to ensure the consistency of the energy grid used in the training data. Spectral smoothing is then applied using a Savitzy-Golay filter from scipy \cite{scipy_source}. The smoothing step is done before the interpolation provided that the inputted spectrum is on an evenly spaced energy scale. The cumulative operation on the spectrum is then performed and this spectrum is the input of the model. The trained RF model is an ensemble of 500 individually trained decision trees, and returns the predictions of each decision tree. A simple average of inferred valence values from each tree is taken as the final prediction, although median and mode predictions can be returned as well. The mode prediction is determined by finding the highest count on a histogram with bin widths of 0.2. The mode is determined by finding the center of the highest bin, meaning integer valence predictions will be returned as 0.1 higher than the integer valence (ie a prediction of Cu(0) will have a mode of 0.1 assigned to it, as the bin will range from 0.0 to 0.2).  The standard deviation of these 500 predictions can approximate the model's internal confidence in its prediction, and is visualized in the prediction histogram in Figures 1, 3 and \ref{model_outline}, the last of which illustrates the entirety of the processing steps performed on an input spectrum.

\subsection*{Experimental EELS}
To validate the utility of this model on experimental data, experimental EELS spectra of standard reference samples were measured, including Cu metal, Cu$_2$O and CuO. Cu metal was purchased from Sigma-Aldrich with 99.999\% purity. Cu$_2$O and CuO were purchased from Sigma-Aldrich with 99.99\% purity. The Cu$_2$O sample was measured using a vacuum holder to prevent oxidation. However, the Cu metal sample was not delivered in a vacuum sealed container, and under the assumption that surface oxidation had already occurred, a vacuum holder was not used for this sample. Using the TEAM I microscope, a double-corrected Thermo Fisher Titan  microscope, we acquired monochromated reference data for these samples at roughly 0.2 meV resolution. Data were collected at 300kV with a semi-convergence angle of 17 mrad and a collection angle of 82 mrad. All data was collected using a Gatan Continuum spectrometer equipped with a K3-IS detector operated in electron counting mode. Spectra were baseline subtracted using the GMS Digital Micrograph software package, and spectra were taken using dual EELS to dynamically remove shifts in the reference elastic energy and deconvolved with the simultaneously measured zero-loss region to mitigate artifacts from electrons experiencing multiple scattering events. The deconvolution of multiple scattering is essential to ensure the experimental EELS spectra are comparable to XAS. 

\section*{Data and Code Availability}
The spectral dataset and the code to generate and analyze the random forest model presented in this study can be found in the GitHub repository https://github.com/smglsn12/ML\_XAS\_EELS. Due to the unpublished nature of this work, this repository is currently private, but will be shared upon request. This repository will be made public upon publication of this work.   

\section*{Author Contributions} 
SPG took the experimental data, generated the simulated XAS dataset, conducted the machine learning training and analysis and wrote the manuscript. DL provided training and expertise necessary to generate the simulated dataset. JC provided experimental EELS knowledge, microscope training, led the collaboration and designed the scope of this work. All authors read, edited and approved the final manuscript.

\section*{Acknowledgment}
This work was primarily funded by the US Department of Energy in the program “4D Camera Distillery: From Massive Electron Microscopy Scattering Data to Useful Information with AI/ML.” Work at the Molecular Foundry was supported by the Office of Science, Office of Basic Energy Sciences, of the U.S. Department of Energy under Contract No. DE-AC02-05CH11231. This research used Theory and Computation resources of the Center for Functional Nanomaterials (CFN), which is a U.S. Department of Energy Office of Science User Facility, at Brookhaven National Laboratory under Contract No. DE-SC0012704. The training and instrumentation support necessary to operate the TEAM I microscope and acquire the experimental EELS data presented in this work was provided by Dr Chengyu Song. 

\bibliography{references}    

\end{document}


\title[Supporting Information For Prediction of the Cu Oxidation State from EELS and XAS Spectra Using Supervised Machine Learning]{Supporting Information For Prediction of the Cu Oxidation State from EELS and XAS Spectra Using Supervised Machine Learning}

\author*[1,2]{\fnm{Samuel P.} \sur{Gleason}}\email{smglsn12@berkeley.edu}

\author[3]{\fnm{Deyu} \sur{Lu}}

\author*[1]{\fnm{Jim} \sur{Ciston}}\email{jciston@lbl.gov}

\affil[1]{\orgdiv{National Center for Electron Microscopy Facility, Molecular Foundry}, \orgname{Lawrence Berkeley National Laboratory}, \orgaddress{\city{Berkeley}, \state{CA}, \country{USA}}}

\affil[2]{\orgdiv{Department of Chemistry}, \orgname{University of California}, \orgaddress{\city{Berkeley}, \postcode{94720}, \state{CA}, \country{USA}}}

\affil[3]{\orgdiv{Center for Functional Nanomaterials}, \orgname{Brookhaven National Laboratory}, \orgaddress{\street{Street}, \city{Upton}, \postcode{100190}, \state{NY}, \country{USA}}}

\maketitle

\setcounter{figure}{0}
\renewcommand{\figurename}{Fig.}
\renewcommand{\thefigure}{S\arabic{figure}}
\begin{figure*}[ht]
    \centering
    \includegraphics[width=\textwidth]{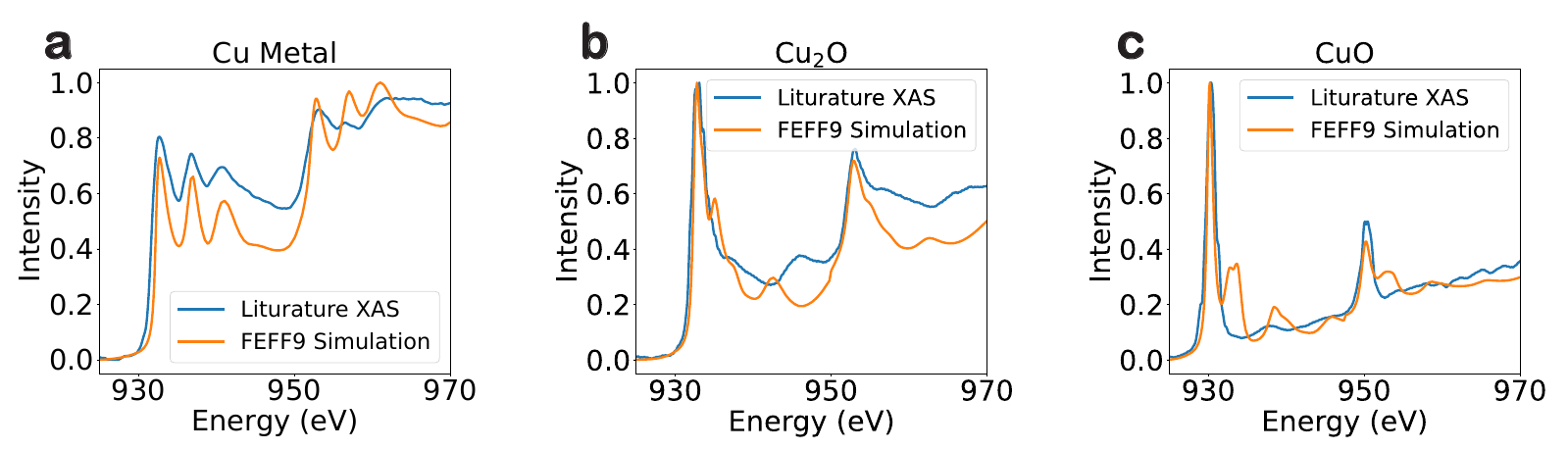}
    \caption{Comparison of the simulated FEFF9 spectra to literature XAS spectra extracted from \cite{xas_paper}. All three comparisons show good qualitative agreement between the experimental and simulated Cu XAS spectra and have been shifted on their energy axis to allow the peaks to align correctly. The Cu(0) spectrum (a) was shifted by -1.0 eV while the other two (b,c) were shifted by -1.2 eV}
    \label{feff_to_lit} 
\end{figure*}

\begin{figure*}[ht]
    \centering
    \includegraphics[width=0.5\textwidth]{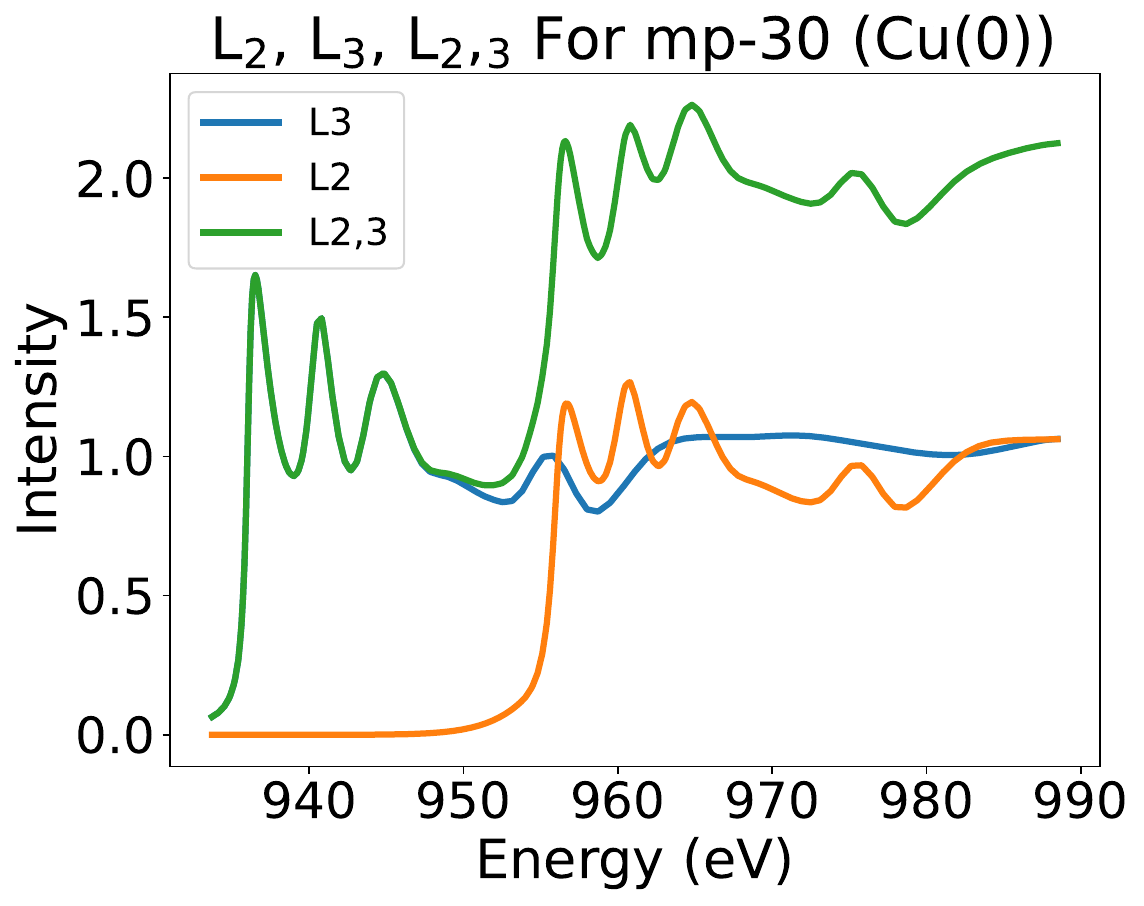}
    \caption{Illustration of how L$_{2,3}$ spectra are created from the individual L$_2$ and L$_3$ spectra. Post interpolation to place the L$_2$ and L$_3$ spectra on the same 0.1 eV energy axis, the two spectra are simply summed together to produce the L$_{2,3}$ spectrum.}
    \label{L2_3} 
\end{figure*}

\begin{figure*}[ht]
    \centering
    \includegraphics[width=0.5\textwidth]{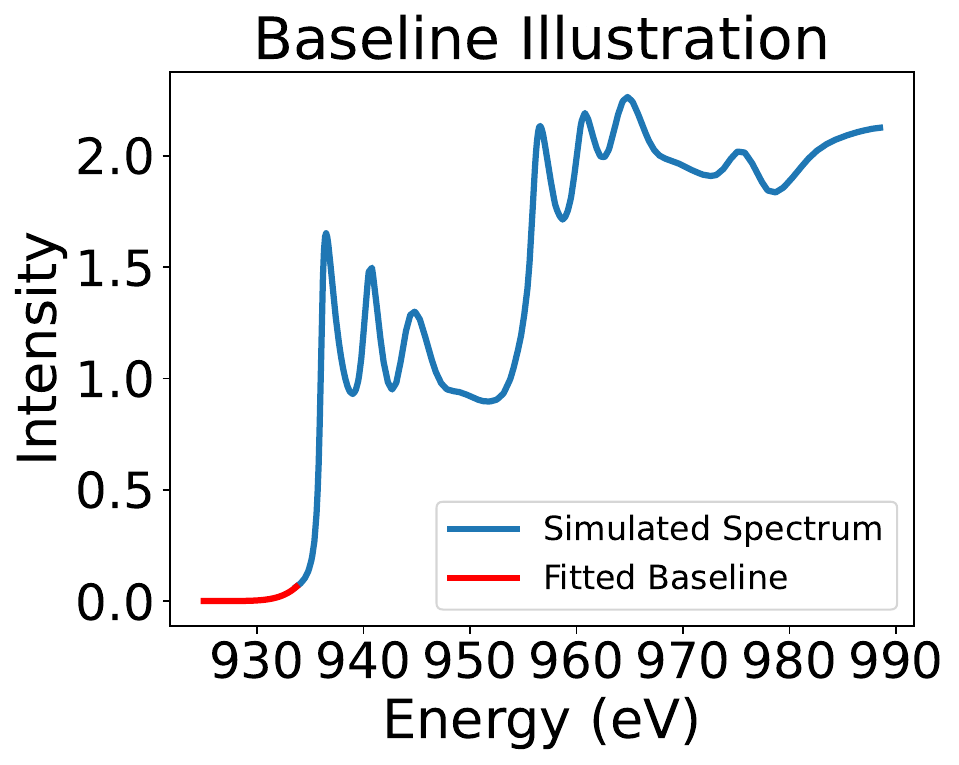}
    \caption{How the pre-edge of each spectrum is scaled so all spectra start at the same energy and have all their intensities start at zero. The final result is a set of spectra running from 925-970 eV with their first values at [925, 0].}
    \label{baseline} 
\end{figure*}

\begin{figure*}[ht]
    \centering
    \includegraphics[width=\textwidth]{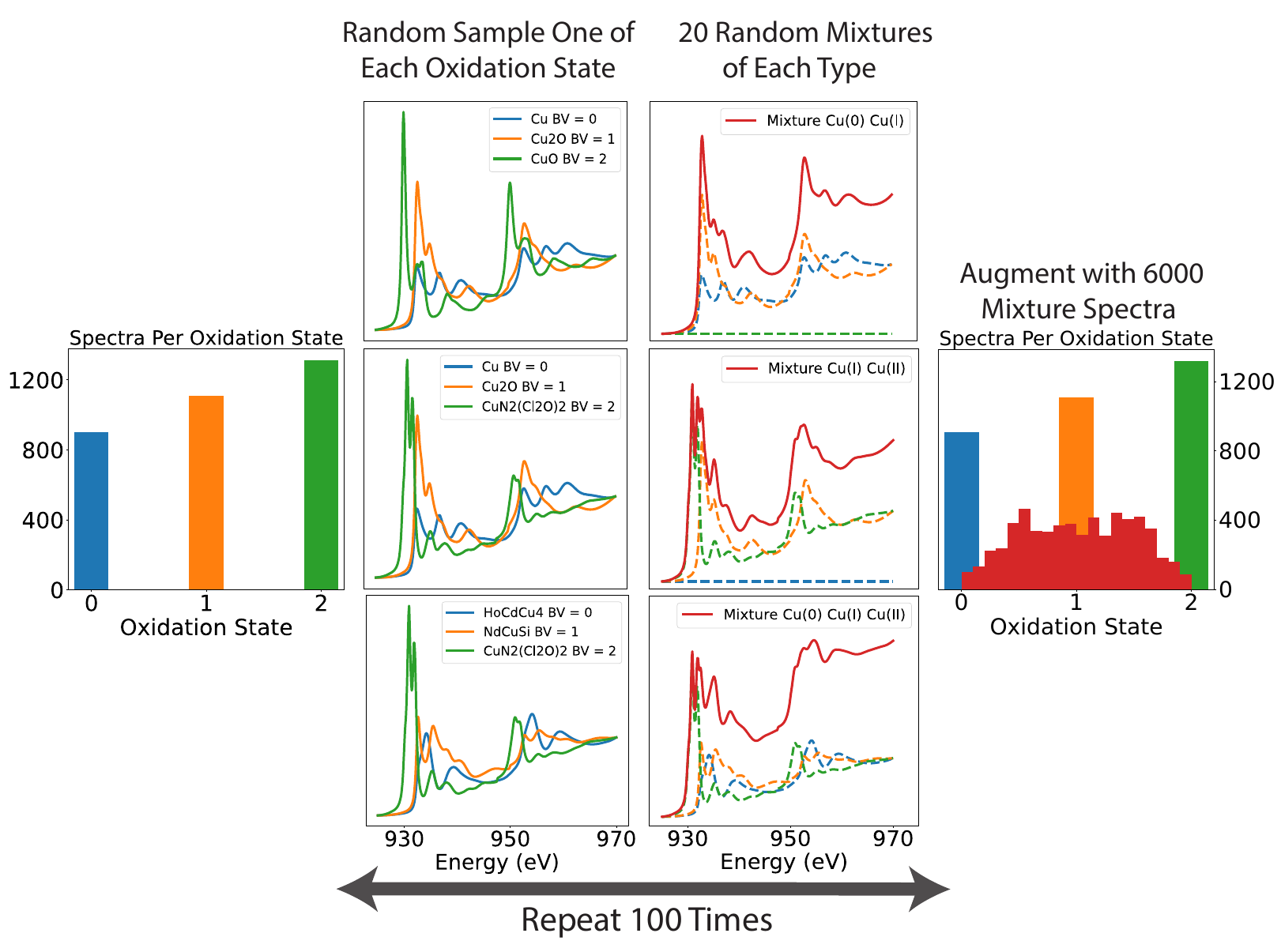}
    \caption{Illustration of data augmentation with simulated mixed valence samples. 100 random spectra are sampled from the full dataset (a,b), and then 20 combinations of Cu(0) and Cu(I) are computed for randomly generated mixture values (c). This process is then repeated two additional times for mixtures of Cu(I) and Cu(II) (d,e), and mixtures of Cu(0), Cu(I) and Cu(II) (f,g). This produces a set of 6000 mixture spectra which fully fills out the spaces between the integer spectra values (h).}
    \label{mixture_generation} 
\end{figure*}

\begin{figure*}[ht]
    \centering
    \includegraphics[width=\textwidth]{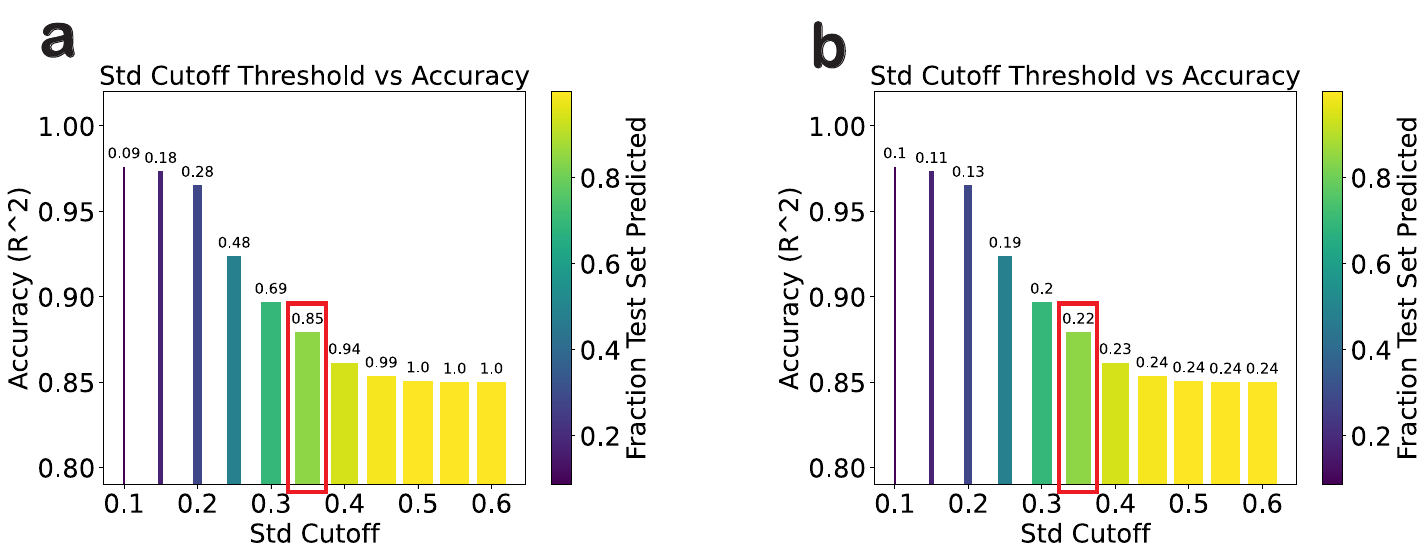}
    \caption{How using specific values of the standard deviation as a threshold to filter out predictions that are highly uncertain changes the percentage of the test set retained (a text above bars) and the RMSE of the retained test set (b text above bars). The red rectangle shows a threshold value of 0.35, where 85\% of the test set is retained. 15\% is discarded because those data points have standard deviations higher than 0.35. When the remaining 85\% of the test set is evaluated, the RMSE falls from 0.24 to 0.22. The color and width of the bars corresponds to the percentage of the test set predicted when that standard deviation threshold is selected. The height of the bars corresponds to the $R^2$ of that standard deviation threshold.}
    \label{std_threshold} 
\end{figure*}

\begin{figure*}[ht]
    \centering
    \includegraphics[width=\textwidth]{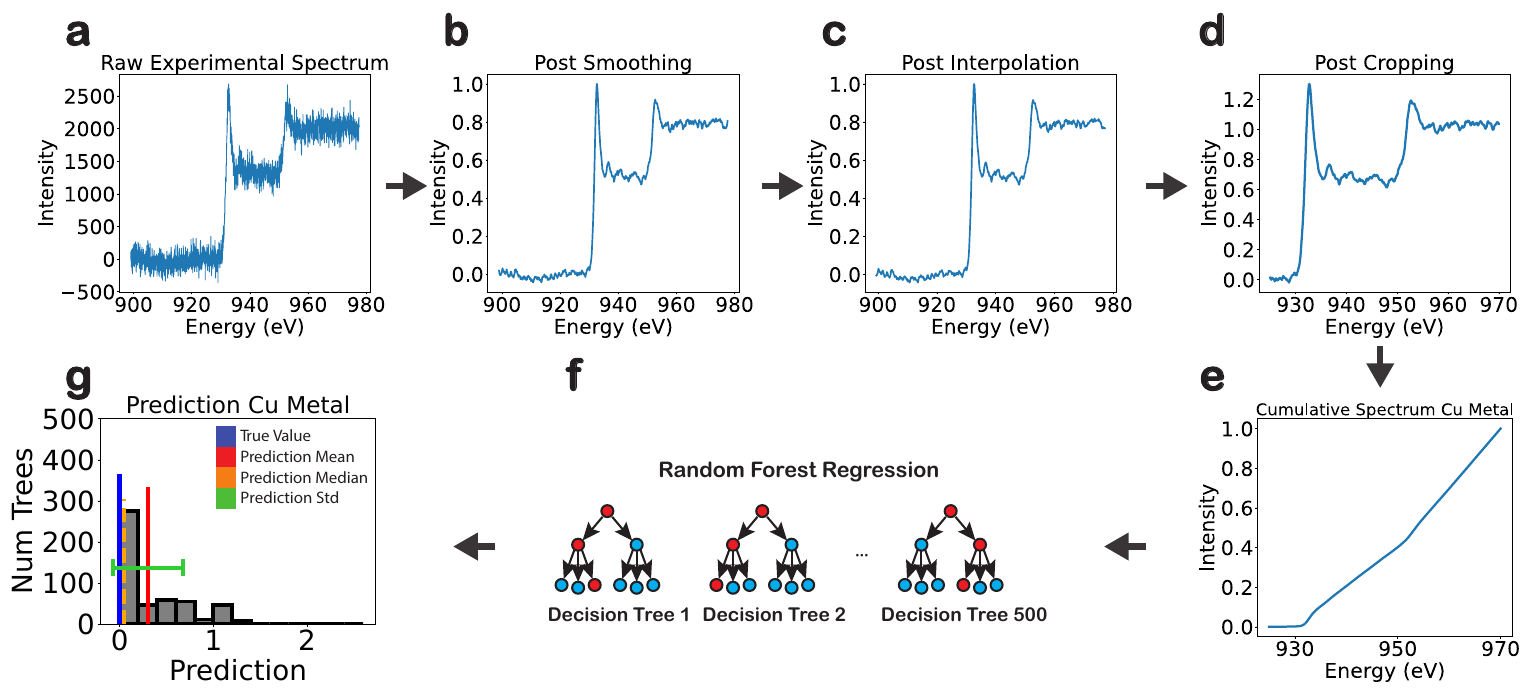}
    \caption{An outline of every processing step for a raw experimental spectrum predicted by this model. After the raw spectrum is taken (a) the spectrum is smoothed with a Savitzky-Golay filter (b). The spectrum is then interpolated onto a 0.1 eV scale (c) and cropped to run from an energy axis of 925-970 eV (d). The spectrum is transformed into a cumulative spectrum (e) and inputted into the random forest model (f). This results in the outputted prediction (g), which includes the prediction determined by the mean (red vertical line) and median (orange dashed line) of the decision trees in the random forest, full predictions across all 500 decision trees (grey bars) and standard deviation (green horizontal line).}
    \label{model_outline} 
\end{figure*}

\begin{figure*}[ht]
    \centering
    \includegraphics[width=\textwidth]{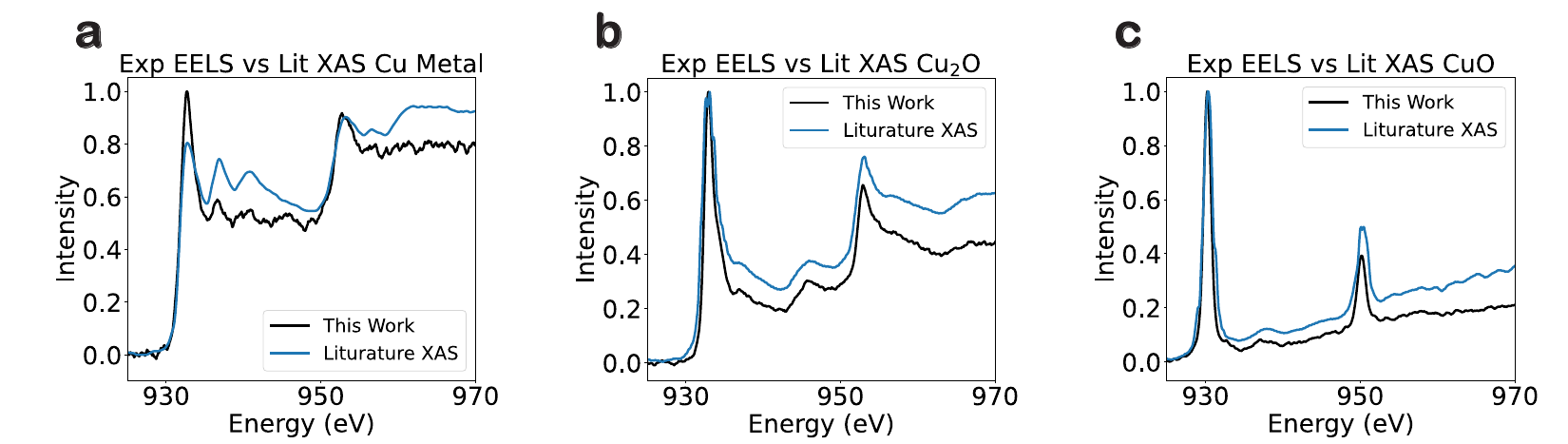}
    \caption{Comparison of Cu(0) (a), Cu(I) (b) and Cu(II) (c) from \cite{xas_paper} and our home institution. The sharp first peak in (a) and its higher intensity relative to the higher energy peaks clearly indicates the presence of Cu(I), as shown in (b).}
    \label{xas_vs_eels} 
\end{figure*}

\begin{figure*}[ht]
    \centering
    \includegraphics[width=\textwidth]{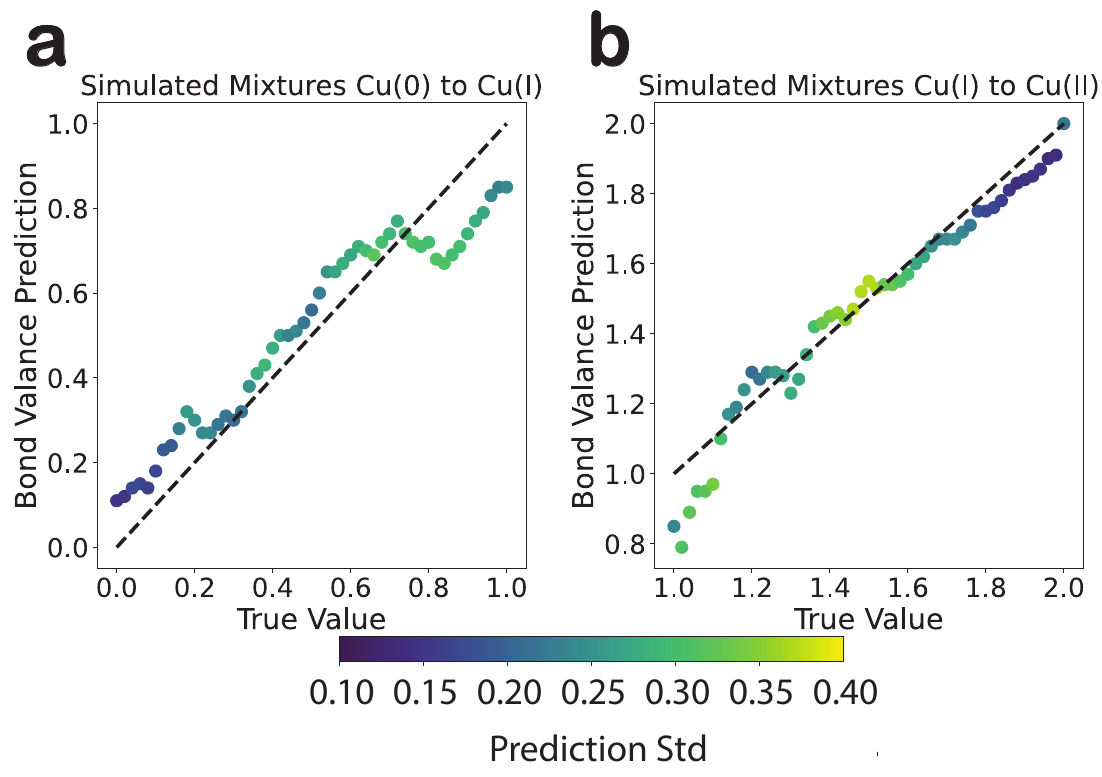}
    \caption{Accuracy of the RF model on simulated mixture test spectra. (a) corresponds to mixtures of Cu(0) and Cu(I) and (b) shows mixtures of Cu(I) and Cu(II). The spectra selected for this visualization are mixtures of Cu(0), Cu$_2$O and CuO. The color of the dots corresponds to the prediction standard deviation. The dashed line indicates the location of a perfect prediction.}
    \label{simulated_mixtures} 
\end{figure*}

\begin{figure*}[ht]
    \centering
    \includegraphics[width=\textwidth]{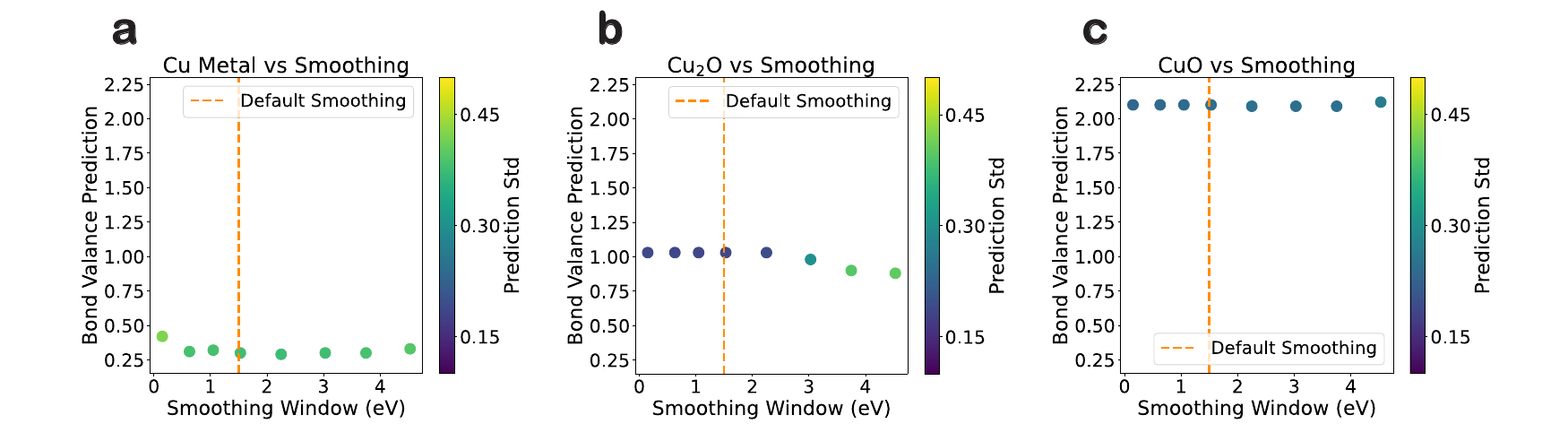}
    \caption{How various amounts of smoothing our experimental standard samples impacts the prediction for Cu(0) (a), Cu(I) (b), and Cu(II) (c). The vertical line indicates the default smoothing condition, window size of 1.5 eV.}
    \label{smoothing_impact} 
\end{figure*}

\begin{figure*}[ht]
    \centering
    \includegraphics[width=\textwidth]{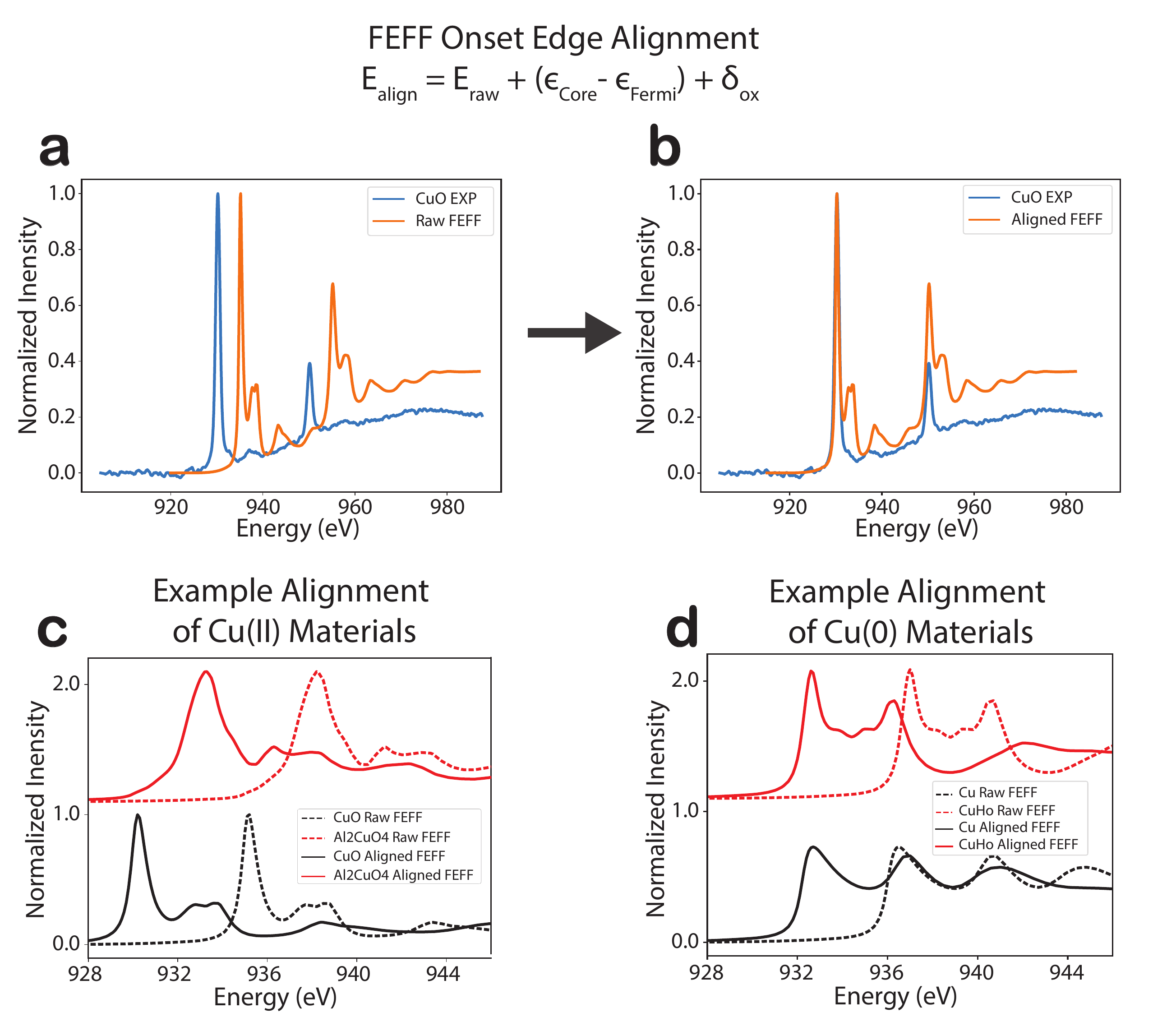}
    \caption{An illustration of the onset edge alignment procedure. The top equation shows the overall alignment equation, where $E_{align}$ is the aligned spectrum and $E_{abs}$ is the unaligned spectrum. $\epsilon_{Fermi}$ is FEFF's estimation of the Fermi energy which is different for each spectrum. $\epsilon_{Core}$ is the estimation of the core hole energy of a 2p core hole in Cu, which is constant for all our spectra. $\delta_{ox}$ is the experimental reference parameter, which is a combination of three parameters whose sum is approximated to be constant for each oxidation state (see methods section). $\delta_{ox}$ ensures the simulated spectra are aligned to experimental samples. There is a different $\delta_{ox}$ for each of the three oxidation states of Cu: 0, +1 and +2.}
    \label{energy_alignment} 
\end{figure*}

\begin{figure*}[ht]
    \centering
    \includegraphics[width=\textwidth]{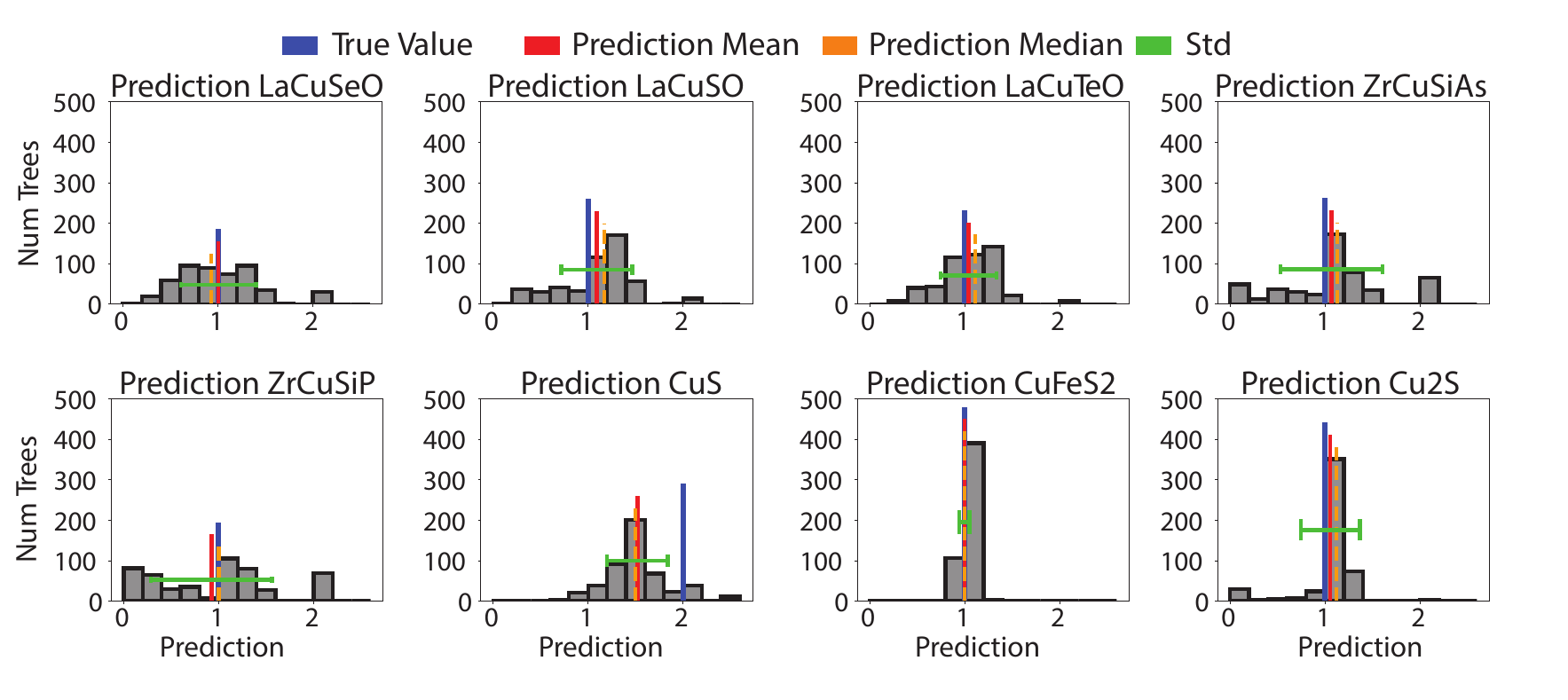}
    \caption{Predictions on eight additional XAS spectra extracted from the literature. The predictions shown here are done with the manual edge alignment shown in Figure \ref{shift_other_exp_samples}. All spectra besides CuS correspond to formal oxidation states of Cu(I) while CuS is Cu(II). The formal oxidation state is shown by the vertical blue line in each plot, while the prediction generated by the mean and median of the decision tree predictions are shown in red and orange, respectively. The standard deviation of the decision tree's predictions is shown as a green horizontal line.}
    \label{other_exp_samples} 
\end{figure*}

\begin{figure*}[ht]
    \centering    \includegraphics[width=\textwidth]{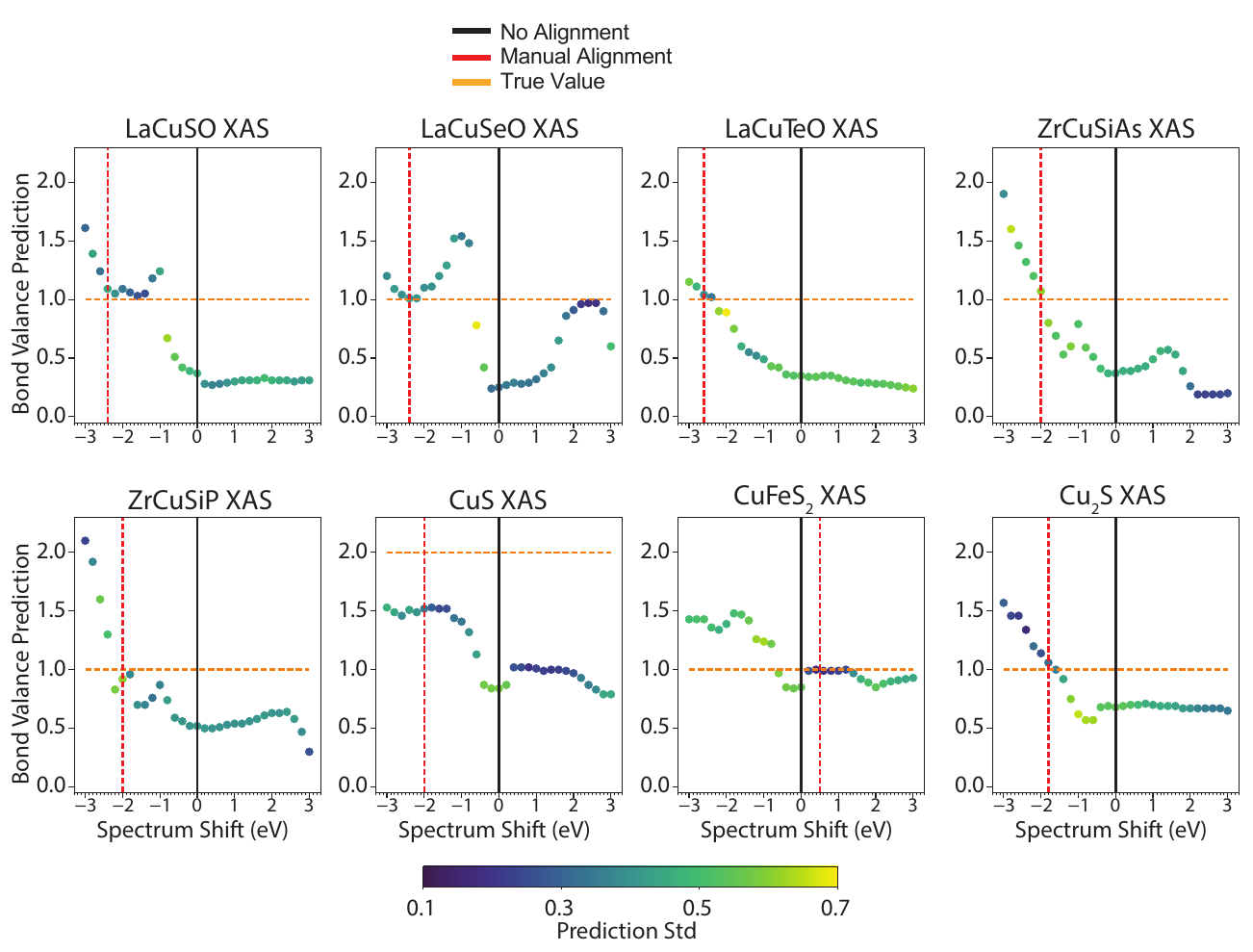}
    \caption{Predictions on eight additional XAS spectra extracted from the literature.  Predictions are shown every 0.2 eV of spectrum shift. Each plot shows the impact of shifting the spectrum’s energy axis on the predicted oxidation state. The solid black line indicates when the raw extracted spectrum is predicted without any additional alignment. The dashed red line indicates when the spectrum has been manually aligned so its L$_3$ onset edge matches its corresponding simulated spectrum in our dataset. All spectra besides CuS correspond to formal oxidation states of Cu(I) while CuS is Cu(II). The formal oxidation state is shown by the horizontal orange line in each plot, while the prediction generated by the mean of the decision tree predictions are shown as scatter spots in each plot, with the prediction standard deviation determined by the color of the spot. 
}
    \label{shift_other_exp_samples} 
\end{figure*}

\begin{figure*}[ht]
    \centering
    \includegraphics[width=\textwidth]{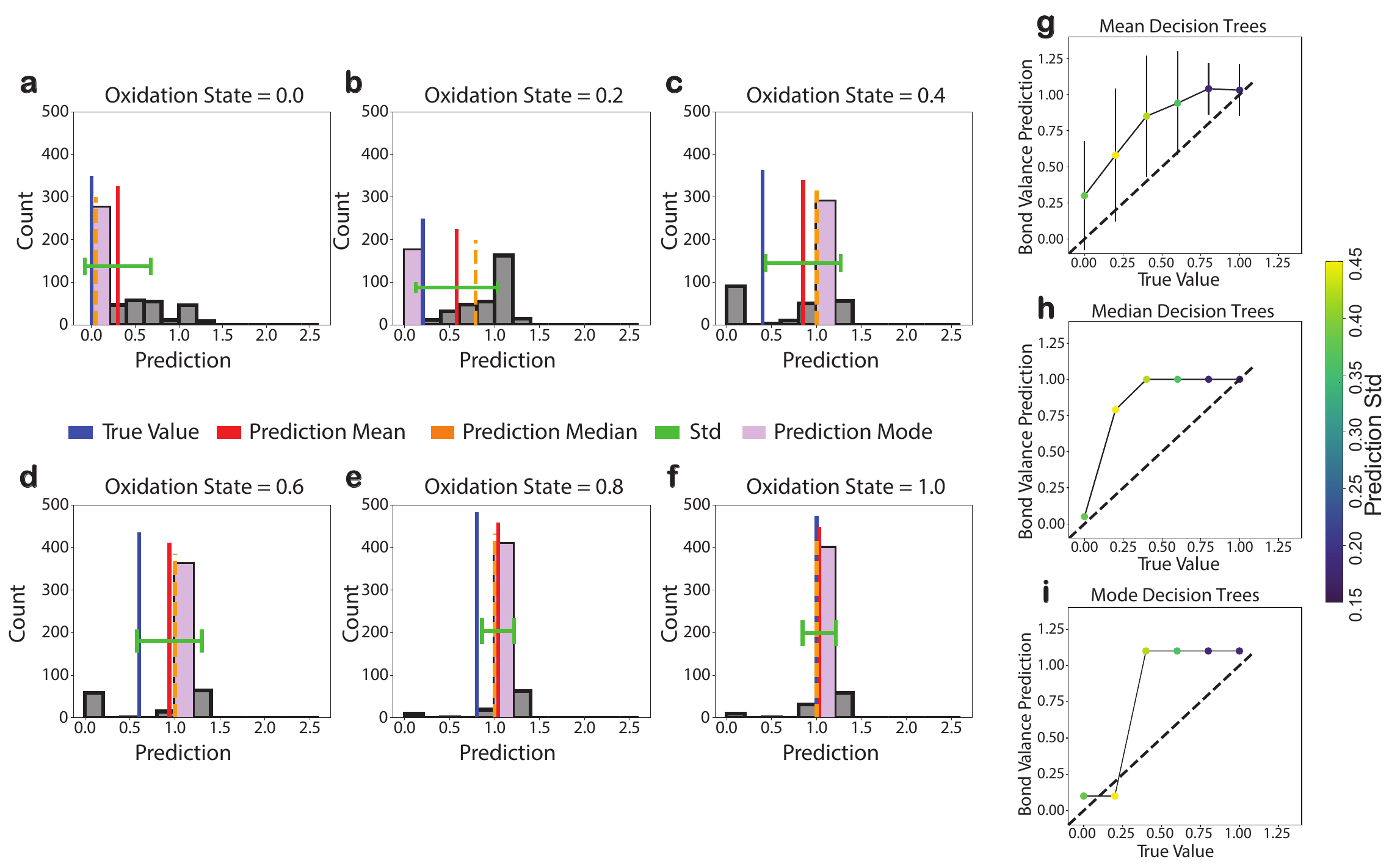}
    \caption{A series of predictions on mixed valent spectra, ranging from pure Cu(0) to pure Cu(I) in 0.2 mixture increments. The following are ratios of Cu(0)/Cu$_2$O: (a) 100/0, (b) 80/20, (c) 60/40, (d) 40/60, (e) 20/80, (f) 0/100. The summary prediction values for using the mean (g) median (h) and mode (i) of the decision trees are shown on the right. The predictions using the mean decision trees (g) have error bars indicating the prediction standard deviation. These are not included in the median and mode predictions (h-i). 
}
    \label{mean_median_and_mode} 
\end{figure*}

\begin{figure*}[ht]
    \centering
    \includegraphics[width=\textwidth]{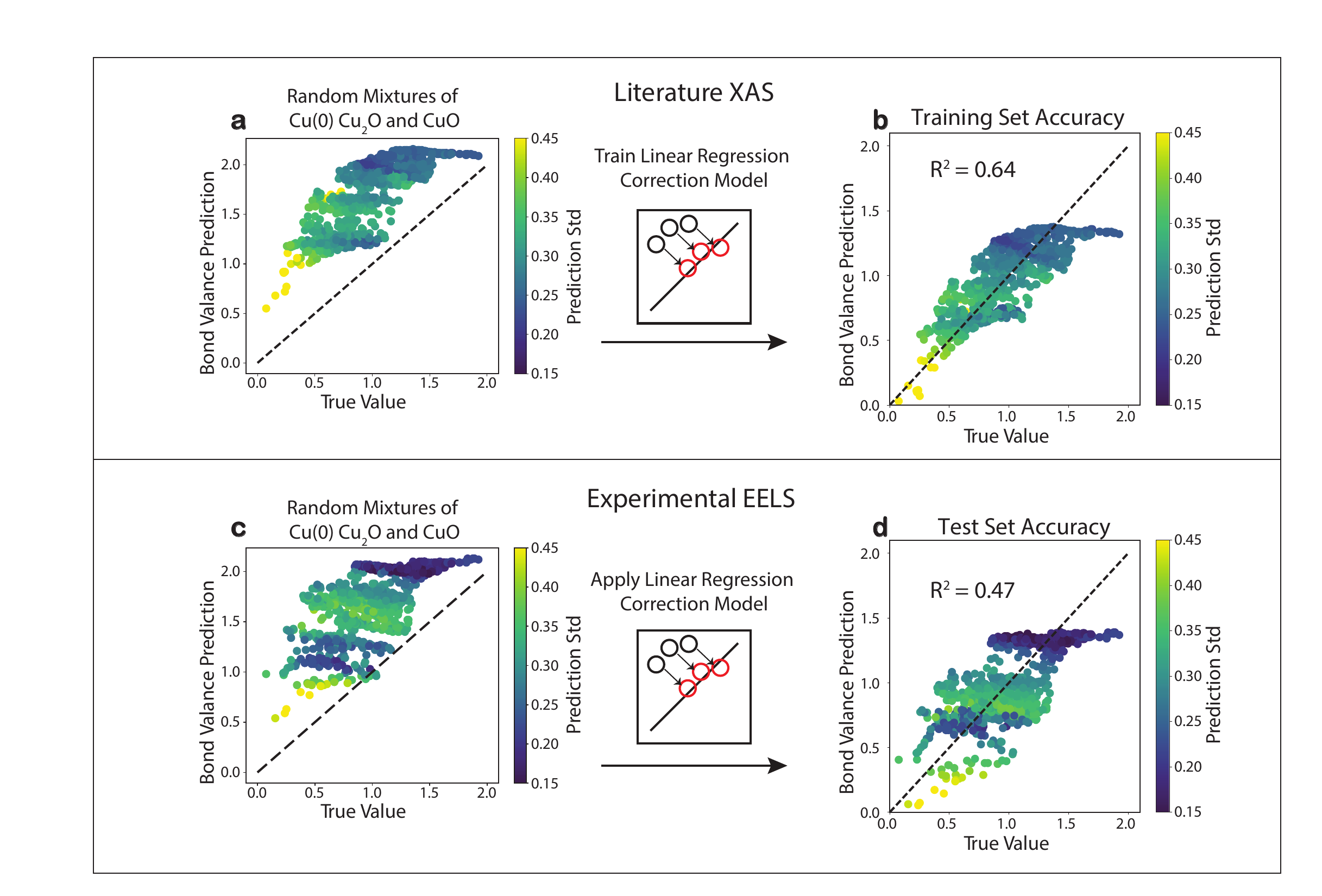}
    \caption{Our random forest model’s predictions of random mixtures of Cu(0), Cu$_2$O and CuO. (a-b) a linear regression model that takes in the mean, median and standard deviation of the RF model’s decision trees and attempts to correct the systematic overestimation. This model is trained on the results in (a) and its training data accuracy is shown in (b). We then demonstrate the utility of this correction by predicting mixtures of experimental EELS spectra containing random proportions of Cu(0), Cu$_2$O and CuO (c) and applying this linear regression model to them (d).}
    \label{empirical_correction} 
\end{figure*}

\begin{figure*}[ht]
    \centering
    \includegraphics[width=\textwidth]{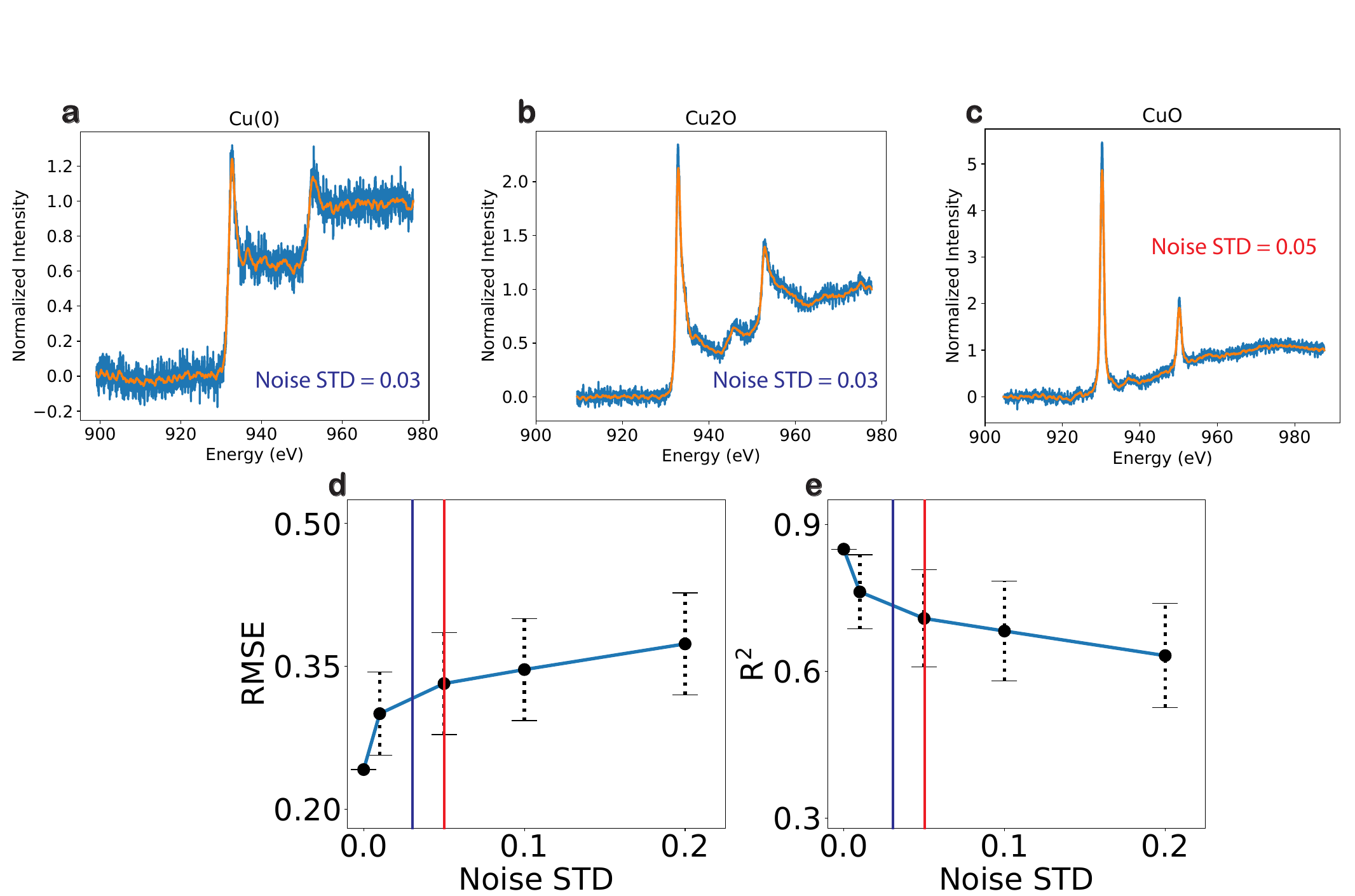}
    \caption{(a-c) Visualization of the noise level of the Cu EELS experimental spectra. (d-e) Where this noise level falls in Figure 6. The blue line corresponds to 0.03 represented in (a-b) while the red line corresponds to 0.05 represented in (c).}
    \label{quantify_noise} 
\end{figure*}

\begin{figure*}[ht]
    \centering
    \includegraphics[width=\textwidth]{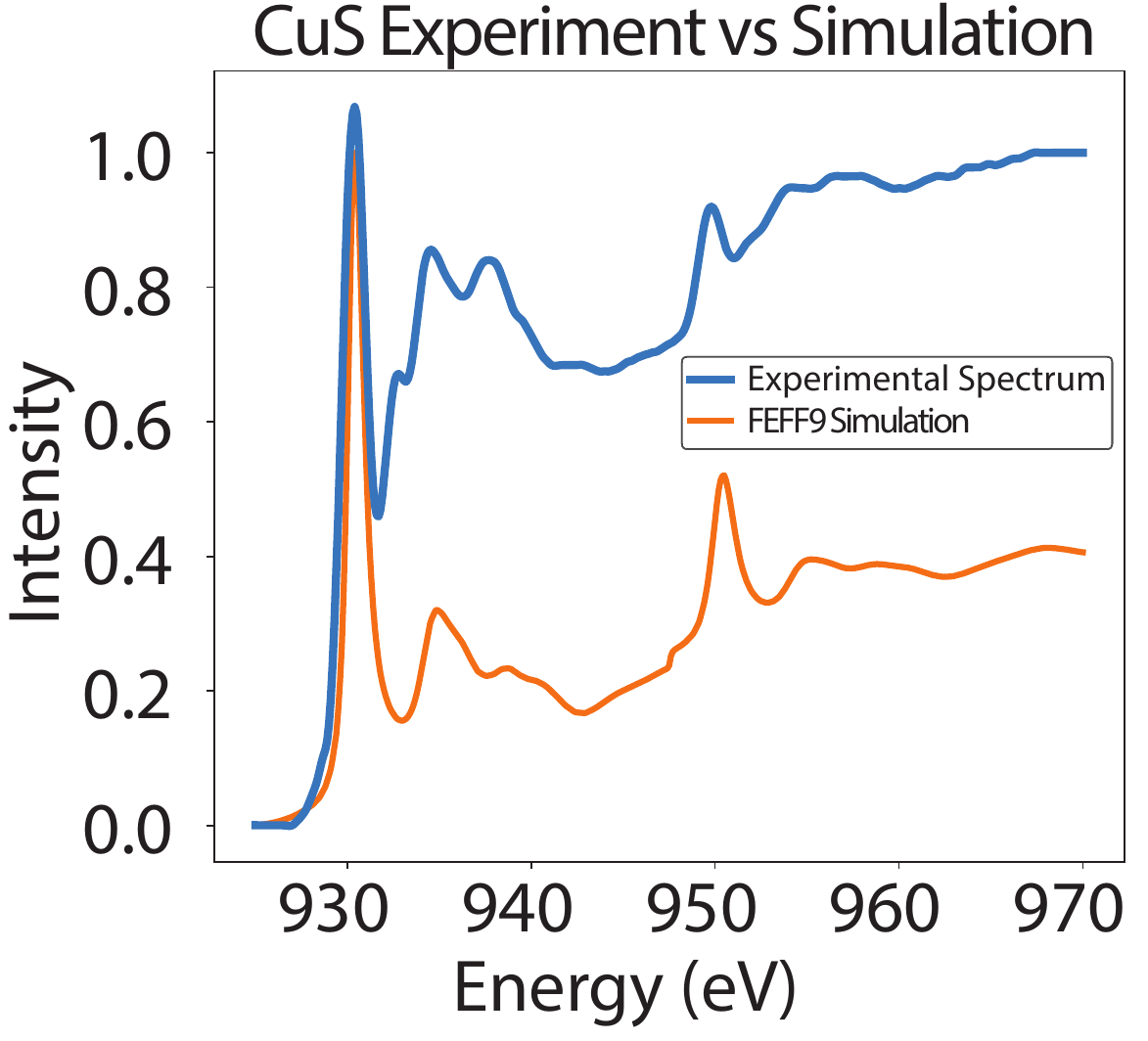}
    \caption{Comparison of the CuS literature XAS spectrum vs our simulated spectrum. The experimental spectrum shows a considerable increase in intensity in the post L$_3$ region, which is indicative of multiple scattering of photons in a thick sample.}
    \label{CuS} 
\end{figure*}

\bibliography{references}